\pdfoutput=1
\documentclass[11pt,twoside,a4paper,cmspaper,final,collab]{cms-tdr}
\def\svnVersion{1e2be67}\def\svnDate{2023/07/19}\def\cmsCernNoTag{CERN-EP-2022-032}\def\cmsCernDate{\today}\def\cmsMessage{Published in Physical Review D as \href{http://dx.doi.org/10.1103/PhysRevD.108.012011}{\doi{10.1103/PhysRevD.108.012011}.}}
\begin{document}\cmsNoteHeader{SUS-21-001}

\newlength\cmsFigWidth
\ifthenelse{\boolean{cms@external}}{\setlength\cmsFigWidth{0.49\textwidth}}{\setlength\cmsFigWidth{0.65\textwidth}}
\ifthenelse{\boolean{cms@external}}{\providecommand{\cmsTable}[1]{#1}}{\providecommand{\cmsTable}[1]{\resizebox{\textwidth}{!}{#1}}}
\newlength\cmsTabSkip\setlength{\cmsTabSkip}{1ex}
\newcommand{\staul}{\ensuremath{\PSGt_{\mathrm{L}}}\xspace}
\newcommand{\staur}{\ensuremath{\PSGt_{\mathrm{R}}}\xspace}
\newcommand{\tautau}{\ensuremath{\tauh\tauh}\xspace}
\newcommand{\dxy}{\ensuremath{d_{xy}}\xspace}
\newcommand{\dR}{\ensuremath{\Delta R}\xspace}
\newcommand{\dphill}{\ensuremath{\Delta\phi(\tauh^{(1)},\tauh^{(2)})}\xspace}
\newcommand{\wjets}{\ensuremath{\PW}+jets\xspace}
\newcommand{\dytautau}{\ensuremath{\PZ/\PGg^{*}\to\PGt\PGt}+jets\xspace}
\newcommand{\sumMT}{\ensuremath{\Sigma\mT}\xspace}
\newcommand{\nj}{\ensuremath{N_{\text{j}}}\xspace}
\newcommand{\lsp}{\PSGczDo}
\newcommand{\ctau}{\ensuremath{c\tau_{0}}\xspace}
\newcommand{\met}{\ptmiss}
\newcommand{\taupt}{\ensuremath{\pt^{\PGt_{\mathrm{h},1}}}\xspace}
\newcommand{\tautwopt}{\ensuremath{\pt^{\PGt_{\mathrm{h},2}}}\xspace}

\cmsNoteHeader{SUS-21-001}
\title{Search for direct pair production of supersymmetric partners of \texorpdfstring{\PGt}{tau} leptons in the final state with two hadronically decaying \texorpdfstring{\PGt}{tau} leptons and missing transverse momentum in proton-proton collisions at \texorpdfstring{$\sqrt{s}=13\TeV$}{sqrt(s) = 13 TeV}}

\date{\today}

\abstract{
A search for the direct production of a pair of \PGt sleptons, the supersymmetric partners of \PGt leptons, is presented. Each \PGt slepton is assumed to decay to a \PGt lepton and the lightest supersymmetric particle (LSP), which is assumed to be stable and to not interact in the detector, leading to an imbalance in the total reconstructed transverse momentum. The search is carried out in events identified as containing two \PGt leptons, each decaying to one or more hadrons and a neutrino, and significant transverse momentum imbalance. In addition to scenarios in which the \PGt sleptons decay promptly, the search also addresses scenarios in which the \PGt sleptons have sufficiently long lifetimes to give rise to nonprompt \PGt leptons. The data were collected in proton-proton collisions at a center-of-mass energy of 13\TeV at the CERN LHC with the CMS detector in 2016--2018, and correspond to an integrated luminosity of 138\fbinv. No significant excess is seen with respect to standard model expectations. Upper limits on cross sections for the pair production of \PGt sleptons are obtained in the framework of simplified models. In a scenario in which the \PGt sleptons are superpartners of left-handed \PGt leptons, and each undergoes a prompt decay to a \PGt lepton and a nearly massless LSP, \PGt slepton masses between 115 and 340\GeV are excluded. In a scenario in which the lifetime of the \PGt sleptons corresponds to $\ctau=0.1\unit{mm}$, where $\tau_{0}$ represents the mean proper lifetime of the \PGt slepton, masses between 150 and 220\GeV are excluded.
}

\hypersetup{
pdfauthor={CMS Collaboration},
pdftitle={Search for direct pair production of supersymmetric partners of tau leptons in the final state with two hadronically decaying tau leptons and missing transverse momentum in proton-proton collisions at sqrt(s)=13 TeV},
pdfsubject={CMS},
pdfkeywords={CMS,  SUSY}}

\maketitle 

\section{Introduction}
\label{sec:intro}

Supersymmetry (SUSY)~\cite{Ramond:1971gb,Golfand:1971iw,Neveu:1971rx,Volkov:1972jx,Wess:1973kz,Wess:1974tw,Fayet:1974pd,Nilles:1983ge} is appealing because it could address some of the shortcomings of the standard model (SM) of particle physics. It features superpartners, \ie, new particles with the same quantum numbers as their SM counterparts, with the exception of their spin, which is shifted by half a unit. The superpartner loop contributions in the radiative corrections to the Higgs boson mass could cancel quadratic divergences, thus solving the fine-tuning problem~\cite{Gildener:1976ai,Veltman:1976rt,'tHooft:1979bh,Witten:1981nf}. In SUSY models with $R$-parity conservation~\cite{Farrar:1978xj}, the lightest supersymmetric particle (LSP) is stable and could be a dark matter (DM) candidate~\cite{Goldberg:1983nd,Ellis:1983ew,darkmatter}.

In this paper, we report the results of a search for the \PGt slepton (\sTau), the superpartner of the \PGt lepton, in proton-proton ($\Pp\Pp$) collisions at a center-of-mass energy of 13\TeV, with the CMS detector. Early universe \sTau-neutralino coannihilation models provide a mechanism that can explain the observed DM relic density~\cite{WMAP,Griest:1990kh,Vasquez,King,Battaglia:2001zp,Arnowitt:2008bz}. These models motivate the existence of a light \sTau as the next-to-lightest supersymmetric particle (NLSP), which would lead to an enhanced rate of production of final states with \PGt leptons in collider experiments~\cite{Belanger:2012jn,Arganda:2018hdn,Arganda:2021qgi}. Here, we study events with two \PGt lepton candidates, both undergoing hadronic decays (\tauh), and having significant transverse momentum imbalance resulting from the presence of LSPs and to a lesser extent, neutrinos from the \PGt lepton decays. While previous searches in this final state have largely focused on prompt decays of the parent particles, the present search also addresses scenarios in which the \sTau is long-lived, which can arise in theories of gauge-mediated SUSY breaking (GMSB) that in many cases predict a \sTau as the NLSP~\cite{Evans:2016zau}. This search is the first to target final states in which reconstructed \tauh candidates are identified as having production vertices that are significantly displaced from the primary interaction point as expected for long-lived \sTau decays.

\begin{figure}[htb]
\centering
\includegraphics[width=0.49\textwidth]{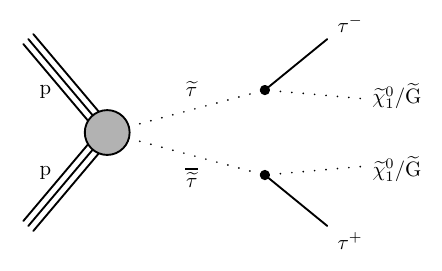}
\caption{\label{fig:diagram} Diagram for direct \sTau pair production, followed by decay of each \sTau to a \PGt lepton and an LSP. For models with promptly decaying \PGt sleptons, the LSP is assumed to be \PSGczDo, the lightest neutralino. For models with long-lived \PGt sleptons, it is assumed to be the gravitino, \sGra.}
\end{figure}

Figure~\ref{fig:diagram} shows a diagram of direct \sTau pair production, with the \sTau decaying to a \PGt lepton and LSP, which we study in this paper within the framework of simplified models~\cite{Alwall:2008ag,Alwall:2008va,Alves:2011wf}. For models featuring prompt decays of the \sTau, we assume \PSGczDo, the lightest neutralino, to be the LSP, and consider a range of \PSGczDo masses up to 200\GeV. We use the symbols \staul and \staur to refer to the superpartners of left- and right-handed \PGt leptons, respectively. We consider cases in which only \staul or only \staur pairs are produced, as well as a degenerate case in which both \staul and \staur pairs are produced. The cross section for \staul pair production is expected to be approximately three times larger than for \staur pair production~\cite{Fuks:2013lya}, while the experimental acceptance is expected to be higher in the right-handed case than in the left-handed because of differences in the polarization of the \PGt leptons that are produced. The small production cross section expected for the signal and the significant SM backgrounds make this search challenging.

Previous searches for direct \sTau pair production in prompt decay scenarios were performed at the CERN LEP collider~\cite{Heister:2001nk,Abdallah:2003xe,Achard:2003ge,Abbiendi:2003ji} and excluded \sTau masses at 95\% confidence level (\CL) up to about 90\GeV for neutralino masses up to 80\GeV, in some models. The ATLAS~\cite{Aad:2014yka,Aad:2015eda} and CMS~\cite{SUS14022} Collaborations performed searches for direct \sTau pair production using 8\TeV CERN LHC data. The ATLAS Collaboration has reported the results of a search for direct \sTau pair production using 13\TeV data corresponding to an integrated luminosity of 139\fbinv~\cite{Aad:2019byo} that excludes \sTau masses between 155 and 310\GeV at 95\% \CL, for the case of a nearly massless LSP. The CMS Collaboration has previously reported the results of a search for direct \sTau pair production using data collected in 2016--2017 at $\sqrt{s}=13\TeV$, corresponding to an integrated luminosity of 77.2\fbinv~\cite{CMS:2019eln}, and placed upper limits on the cross section for \sTau pair production.

We also target scenarios where the \sTau has a short but finite lifetime, decaying within a few\unit{cm} of the primary interaction point. These signatures are sensitive to GMSB SUSY models in which a nearly massless gravitino (\sGra) is the LSP and the \sTau is the NLSP that can become long-lived as a result of its suppressed coupling to the gravitino. We consider models involving the pair production of $\sTau_{1}$, a mixture of \staul and \staur, for \ctau up to 2.5\unit{mm}, where $\tau_0$ is the mean proper lifetime of the $\sTau_{1}$. For these models, we assume a scenario with maximal mixing, \ie, with a mixing angle of $\frac{\pi}{4}$, for which the cross section is very similar to that for purely right-handed \sTau pair production~\cite{Fuks:2013lya}. This scenario is therefore consistent with GMSB SUSY models, in which the \sTau is typically right-handed.  Previously, the LEP experiments set limits on GMSB SUSY models with $\sTau_{1}$ as the NLSP, with the strongest limits coming from the OPAL experiment~\cite{Abbiendi:2005gc}. The OPAL limits excluded masses up to 87.4\GeV at 95\% \CL, for all $\sTau_{1}$ lifetimes. The ATLAS Collaboration recently reported the results of a search for long-lived sleptons, including $\sTau_{1,2}$, a combination of mixed states of \staul and \staur, in final states with nonprompt electrons or muons. The degenerate production of the two mixed states $\sTau_{1}$ and $\sTau_{2}$ results in a larger cross section than the production of $\sTau_{1}$ alone. The search excluded $\sTau_{1,2}$ masses up to 340\GeV for a proper lifetime of 0.1\unit{ns}, \ie, for a \ctau of 30\unit{mm}, within the simplified GMSB SUSY model considered~\cite{Aad:2020bay}, under the assumption of a nearly massless LSP. 

The results presented in this paper supersede those of the search reported in Ref.~\cite{CMS:2019eln}, for final states with two \tauh candidates and missing transverse momentum (\ptmiss), and also include scenarios with long-lived \PGt sleptons. The LHC $\Pp\Pp$ collision data collected with the CMS detector in 2018 have been analyzed, and the data collected in 2016--2017 have been re-analyzed, resulting in a sample corresponding to a total integrated luminosity of 138\fbinv. Improved techniques are used to describe the SM background with \PGt leptons through a method called ``embedding"~\cite{Sirunyan:2019drn}, which estimates the background with two genuine \PGt lepton candidates by selecting dimuon events in data, removing reconstructed muons, and replacing them with simulated \PGt lepton decays. Updated methods are used for the \tauh candidate selection~\cite{CMS:2022prd} and the search region definitions have been re-optimized. Overall, we obtain a significant improvement in the search sensitivity. Tabulated results are provided in the HEPData record for this analysis~\cite{hepdata}.

\section{The CMS detector}
\label{sec:detector}
The central feature of the CMS apparatus is a superconducting solenoid of 6\unit{m} internal diameter, providing a magnetic field of 3.8\unit{T}. A silicon pixel and strip tracker, a lead tungstate crystal electromagnetic calorimeter, and a brass and scintillator hadron calorimeter, each composed of a barrel and two endcap sections, reside within the solenoid volume. Forward calorimeters extend the pseudorapidity ($\eta$) coverage provided by the barrel and endcap detectors. Muons are detected in gas-ionization chambers embedded in the steel flux-return yoke outside the solenoid. Events of interest are selected using a two-tiered trigger system. The first level (L1), composed of custom hardware processors, uses information from the calorimeters and muon detectors to select events at a rate of around 100\unit{kHz} within a fixed latency of about 4\mus~\cite{CMS:2020cmk}. The second level, known as the high-level trigger (HLT), consists of a farm of processors running a version of the full event reconstruction software optimized for fast processing, and reduces the event rate to around 1\unit{kHz} before data storage~\cite{CMS:2016ngn}.

\section{Event reconstruction and simulation}
\label{sec:evtreco}
The event reconstruction uses the particle-flow (PF) algorithm~\cite{Sirunyan:2017ulk}, which aims to reconstruct and identify individual particles in an event, with an optimized combination of information from the various elements of the CMS detector. The vector \ptvecmiss is computed as the negative vector sum of the transverse momenta, \ptvec, of all PF-reconstructed particles in an event. Its magnitude, \ptmiss, is used in the search as a discriminator between signal and SM background. Events selected for the search are required to pass selection criteria~\cite{Sirunyan:2019kia} designed to remove anomalous high-\ptmiss events that can occur due to a variety of reconstruction failures, detector malfunctions, or noncollision backgrounds, and must have at least one reconstructed $\Pp\Pp$ interaction vertex. The primary vertex (PV) is taken to be the vertex corresponding to the hardest scattering in the event, evaluated using tracking information alone, as described in Section 9.4.1 of Ref.~\cite{CMS-TDR-15-02}.

Reconstructed particles are clustered into jets using the anti-\kt algorithm~\cite{Cacciari:2008gp,Cacciari:2011ma} with a distance parameter of 0.4. Jet momentum is determined as the vectorial sum of all particle momenta in the jet, and is found from simulation to be, on average, within 5 to 10\% of the true momentum over the whole \pt spectrum and detector acceptance. Additional $\Pp\Pp$ interactions within the same or nearby bunch crossings (pileup) can contribute additional tracks and calorimetric energy depositions, increasing the apparent jet momentum. To mitigate this effect, tracks identified to be originating from pileup vertices are discarded and an offset correction is applied to correct for remaining contributions. Jet energy corrections are derived from simulation studies so that the average measured energy of jets becomes identical to that of particle level jets. In situ measurements of the momentum balance in dijet, photon+jet, {\PZ}+jet, and multijet events are used to determine any residual differences between the jet energy scale in data and in simulation, and appropriate corrections are made~\cite{CMS:2016lmd}. Jets considered in this analysis, other than those from which \tauh candidates are reconstructed, are required to be within the tracker volume, $\abs{\eta} < 2.4$, and to satisfy the condition $\pt>30\GeV$. They are required to be separated in the plane of $\eta$ and azimuthal angle ($\phi$) by $\dR \equiv \sqrt{\smash[b]{(\Delta\eta)^2 + (\Delta\phi)^2} } > 0.4$ from \tauh candidates to avoid the double counting of objects. The deep neural network (DNN) based combined secondary vertex algorithm (\textsc{DeepCSV})~\cite{Sirunyan:2017ezt} is used to identify, or ``tag", jets originating from the hadronization of {\cPqb} quarks. A high-efficiency (``Loose") working point of this algorithm is used to reject events with {\cPqb} quark jets that are likely to have originated from SM backgrounds with top quarks. This working point corresponds to an efficiency of $\approx$84\% for identifying {\cPqb} quarks originating from top quark decays, and misidentification rates of about 41 and 11\%, respectively, for jets from charm quarks and from light quarks or gluons.

In order to suppress SM backgrounds, such as those originating from diboson production, or \ttbar production in association with a vector boson, we veto events with isolated electron or muon candidates. We identify these charged leptons using the same selection criteria as those described in~\cite{CMS:2019eln}. The \tauh candidates are reconstructed from jets, using the hadrons-plus-strips algorithm~\cite{Sirunyan:2018pgf}, which combines 1 or 3 tracks with energy deposits in the calorimeters, to identify the \tauh decay modes. Decay modes with one or three charged hadrons, with or without neutral pions, are considered in this search. To distinguish genuine \tauh decays from electrons, muons, or jets originating from the hadronization of quarks or gluons, the multi-class DNN-based \textsc{DeepTau}~\cite{CMS:2022prd} algorithm is used. Information from all individual reconstructed particles near the \tauh axis is combined with properties of the \tauh candidate and event activity. We employ both a relaxed (``Loose") and a more stringent (``VVTight") working point of the anti-jet \textsc{DeepTau} discriminant. These have efficiencies of $\approx$80 and $\approx$40\% for a genuine \tauh decay in the case of prompt \tauh production, respectively, and misidentification rates of $\approx$0.5 and $\approx$0.06\%, respectively, for quark and gluon jets. Since the \textsc{DeepTau} algorithm was optimized for promptly produced \tauh candidates, the efficiency is reduced in the case of \tauh candidates originating from long-lived \PGt slepton decays. The efficiency of the ``VVTight" working point, which is used to select signal \tauh candidates, ranges between 20\% and 30\% for the largest \sTau lifetimes considered in the search regions that explicitly target displaced \tauh decays.

Monte Carlo (MC) simulation is used to model the signal. The \MGvATNLO version 2.3.3 and 2.4.2 event generators~\cite{Alwall:2014hca} are used at leading order (LO) precision to generate models of direct \sTau pair production for promptly decaying \sTau up to the production of \PGt leptons, for simulated event samples corresponding to the 2016 and 2017--2018 data sets, respectively. The \PGt lepton decays are modeled by \PYTHIA~8.2 (8.212 for 2016, or 8.230 for 2017--2018)~\cite{Sjostrand:2014zea}. For signal models in which the \sTau undergoes nonprompt decays, the \sTau pair production process is generated with \MGvATNLO at LO precision, with the \sTau decays being subsequently carried out by \PYTHIA, using a specified \sTau lifetime. We consider models with \ctau ranging from 0.01 to 2.5\unit{mm}. In the case of both prompt and long-lived \PGt sleptons, we consider \sTau masses ranging from 90 to 500\GeV. The CUETP8M1 underlying event tune~\cite{Khachatryan:2015pea} is used with \PYTHIA for simulated event samples corresponding to the 2016 data set, and the CP5 tune~\cite{Sirunyan:2019dfx} is used for 2017 and 2018 samples. The NNPDF3.0LO~\cite{Ball:2014uwa} set of parton distribution functions (PDFs) is used in generating the 2016 simulation samples, while the NNPDF3.1 next-to-leading order (NLO) PDFs are used for 2017--2018. Showering and hadronization of partons are carried out using \PYTHIA, while a detailed simulation  of the CMS detector is based on the \GEANTfour~\cite{geant4} package. Finally, uncertainties in the renormalization and factorization scales have been obtained using the \textsc{SysCalc} package~\cite{Kalogeropoulos:2018cke}. The signal production cross sections are calculated at NLO using next-to-leading logarithmic (NLL) soft-gluon resummation~\cite{Beenakker:1999xh,Fuks:2013lya}.

We also use MC simulation to model the background from SM production of Higgs bosons. The {\POWHEG}v2~\cite{Nason:2004rx,Frixione:2007vw,Alioli:2010xd,Re:2010bp} generator is used to produce samples of Higgs boson events with decays to \PGt lepton pairs. Other backgrounds, originating from processes that give rise to two genuine \tauh decays, or to one or more jets that are misidentified as \tauh candidates, are estimated from data, as described in Section~\ref{sec:bkgest}. The background from events in which an electron or muon is misidentified as a \tauh candidate is found from simulation to be negligible.

Simulated events are weighted to match the pileup profile observed in data. Scale factors are applied to simulated events to account for differences with respect to data in trigger efficiencies, \tauh identification efficiencies, jet and \tauh energy scales, and {\cPqb} tagging efficiency. We improve the modeling of initial-state radiation (ISR) in the 2016 signal simulation samples by reweighting the $\pt^{\mathrm{ISR}}$ distribution, where $\pt^{\mathrm{ISR}}$ corresponds to the magnitude of the total transverse momentum of the system of parent SUSY particles, obtained in our case by calculating the vector sum of the \pt of the two \PGt sleptons at generator level. This reweighting procedure is based on studies of the \pt of \PZ bosons in data and simulation~\cite{Chatrchyan:2013xna}. No corrections were found to be necessary for the $\pt^{\mathrm{ISR}}$ distribution in 2017 and 2018 simulation samples, as the ISR modeling was improved in the simulation with the updated underlying-event tune.

\section{Event selection}
\label{sec:evtsel}

The data used in this search are selected with two sets of triggers: a trigger requiring the presence of two \tauh candidates, each with $\pt>35$ ($>$40)\GeV in 2016 (2017--2018) data, and a \ptmiss-based trigger, with a threshold varying between 100 and 140\GeV, depending on the data-taking period. The di-\tauh trigger efficiency reaches its plateau, which ranges between 75\% and 95\% depending on the \tauh decay mode and the data-taking period, for \tauh \pt values above 80\GeV. The di-\tauh trigger is used for events with $\ptmiss<200$\GeV, while the \ptmiss-based trigger is used for events with $\ptmiss>200$\GeV. The offline \ptmiss threshold is chosen to be close to the range of approximately 250--300\GeV in which the efficiency of the \ptmiss-based trigger reaches its plateau of $\sim$95--98\%, depending on the data-taking period.

After the trigger selection, we impose a baseline event selection requiring the presence of exactly two \tauh candidates of opposite charge with $\pt>40\GeV$, $\abs{\eta}<2.1$, and satisfying the ``VVTight" \textsc{DeepTau} selection and other criteria described in Section~\ref{sec:evtreco}. Backgrounds originating from diboson production or \ttbar production in association with a vector boson are suppressed by vetoing events with electron or muon candidates with $\pt>20\GeV$ and $\abs{\eta}<2.5$ or $<$2.4 for electrons and muons respectively, or additional \tauh candidates with $\pt>30\GeV$ satisfying the ``Loose" \textsc{DeepTau} selection. We reject any events with a {\cPqb}-tagged jet to suppress top quark backgrounds, and we require $\abs{\Delta\phi(\tauh^{(1)},\tauh^{(2)})}>1.5$. This requirement retains high signal efficiency while reducing the background from \dytautau events in which the di-\tauh system is boosted, resulting in a smaller angular separation between the \tauh candidates. Finally, we require $\ptmiss>50\GeV$ to suppress the background with two misidentified \tauh candidates. To avoid effects related to jet mismeasurement that can contribute to spurious \ptmiss, we require the \ptvecmiss to have a minimum separation of 0.25 in $\abs{\Delta \phi}$ from reconstructed jets.

The search strategy relies on a simultaneous maximum likelihood fit of the event yields observed in 31 search regions (SRs), which are described below and summarized in Table~\ref{tab:srdefinitions}. We use a number of discriminants to subdivide events satisfying the baseline selection criteria into exclusive SRs. For signal events, we expect the two stable LSPs in the final state to contribute to the \ptmiss. Consequently, we expect the correlations between \ptvecmiss and the reconstructed \tauh candidates to be different between signal and background events. Mass observables calculated from \ptvecmiss and the \ptvec of the \tauh can be used to exploit these differences and discriminate signal from background. 

One of the discriminants used is \sumMT, the sum of the transverse masses (\mT) calculated for each \tauh candidate with \ptmiss, given by
\begin{equation}
\sumMT = \mT(\tauh^{(1)}) + \mT(\tauh^{(2)}),
\end{equation}
where the transverse mass for each \tauh candidate is
\begin{equation}
\mT(\tauh) \equiv \sqrt{2 \pt^{\tauh} \ptmiss [1 - \cos \Delta\phi(\ptvec^{\tauh}, \ptvecmiss)]}.\label{eq:MT}
\end{equation}
For our signal models, \ptmiss can originate from neutrinos from \tauh decays, as well as from LSPs. However, the predominant contribution to the \ptmiss is expected to come from the LSPs, which we assume to be massless in the calculation of \mT.

We also use the ``stransverse mass" \mTii~\cite{MT2variable,MT2variable2, MT2variable3}, given by
\begin{equation}
\mTii = \min_{\ptvec^{\mathrm{X}(1)} + \ptvec^{\mathrm{X}(2)} = \ptvecmiss}
  \left[ \max \left( \mT^{(1)} , \mT^{(2)} \right) \right],
\label{eq:MT2}
\end{equation}
where $\ptvec^{\mathrm{X}(i)}$  (with $i$=1, 2) are the unknown transverse momenta of the two invisible particles, X(1) and X(2), corresponding to the LSPs in our signal models, and $\mT^{(i)}$ are the transverse masses calculated from the assigned LSP transverse momenta. The latter are obtained by associating either of the two LSPs, assumed to be massless, to one of the two parent \sTau decays. The minimization of Eq. (\ref{eq:MT2}) is performed over all possible LSP 2-vector momenta, which are constrained to add up to the \ptvecmiss in the event. We expect large values of \mTii to occur more frequently in signal events for models with larger \PSGt masses, and to occur relatively rarely in SM background events.

For the selection of events in the SRs, requirements of $\mTii>25\GeV$ and $\sumMT>200\GeV$ are imposed. Two sets of SRs are defined: the ``prompt" SRs, targeting models in which the \sTau decays promptly, and the ``displaced" SRs, targeting long-lived \sTau models in which nonprompt \tauh candidates are expected.

In order to ensure that the prompt and displaced SRs are disjoint, we require that events in the prompt SRs have at least one \tauh that does not satisfy the ``displaced {\tauh}" criteria described below for the displaced SRs. Events are then subdivided into bins of \mTii and \sumMT, which provides sensitivity to a range of \sTau masses. We further subdivide events into two categories based on the number of reconstructed jets (\nj): $\nj=0$, and $\nj \geq 1$. Since background events that satisfy the SR kinematic selection criteria usually contain additional jets, the 0--jet category provides SRs with improved signal-to-background ratios. Signal events with ISR or pileup jets may populate the $\nj \geq 1$ SRs, and so we also retain these to avoid losing signal sensitivity. Finally, we gain additional sensitivity in bins with lower \sumMT and \mTii values ($\sumMT<300\GeV$ and $\mTii<75\GeV$) in the 0--jet category, which have relatively high background, by further subdividing them into two bins based on the \pt of the leading (higher-\pt) \tauh candidate, \taupt ($\taupt<90\GeV$, and $\taupt \geq 90\GeV$). This further improves the discrimination of signal from background as \tauh in signal events tend to have higher \pt.

The displaced category is defined by imposing the following ``displaced {\tauh}" criteria for both \tauh candidates. We require the significance of the \tauh impact parameter relative to the PV in the transverse plane (\dxy), defined as the quantity divided by its uncertainty, to have an absolute value above 5, and the absolute value of its three-dimensional impact parameter (IP3D) to exceed 100\micron. We also require $\abs{\dphill}>1.75$ to further suppress the background in the displaced category. For events satisfying these selection criteria, the \pt of the subleading (lower-\pt) \tauh candidate, \tautwopt, provides additional discrimination between signal and the remaining background. Accordingly, we define two SR bins for events in this category, with $\tautwopt<110\GeV$ and $\tautwopt \geq 110\GeV$.

\begin{figure*}[htb!]
\centering
\includegraphics[width=0.48\textwidth]{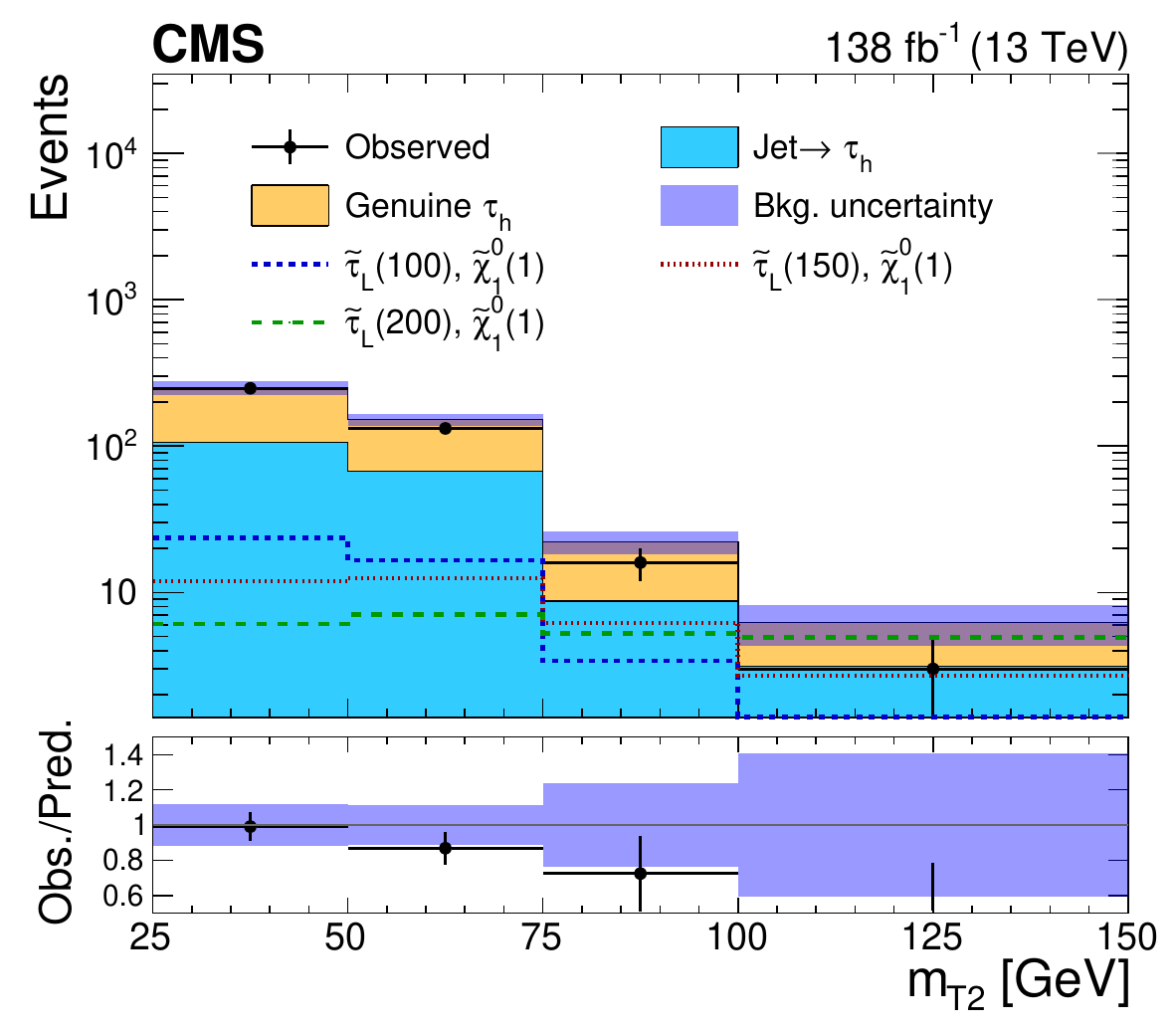}
\includegraphics[width=0.48\textwidth]{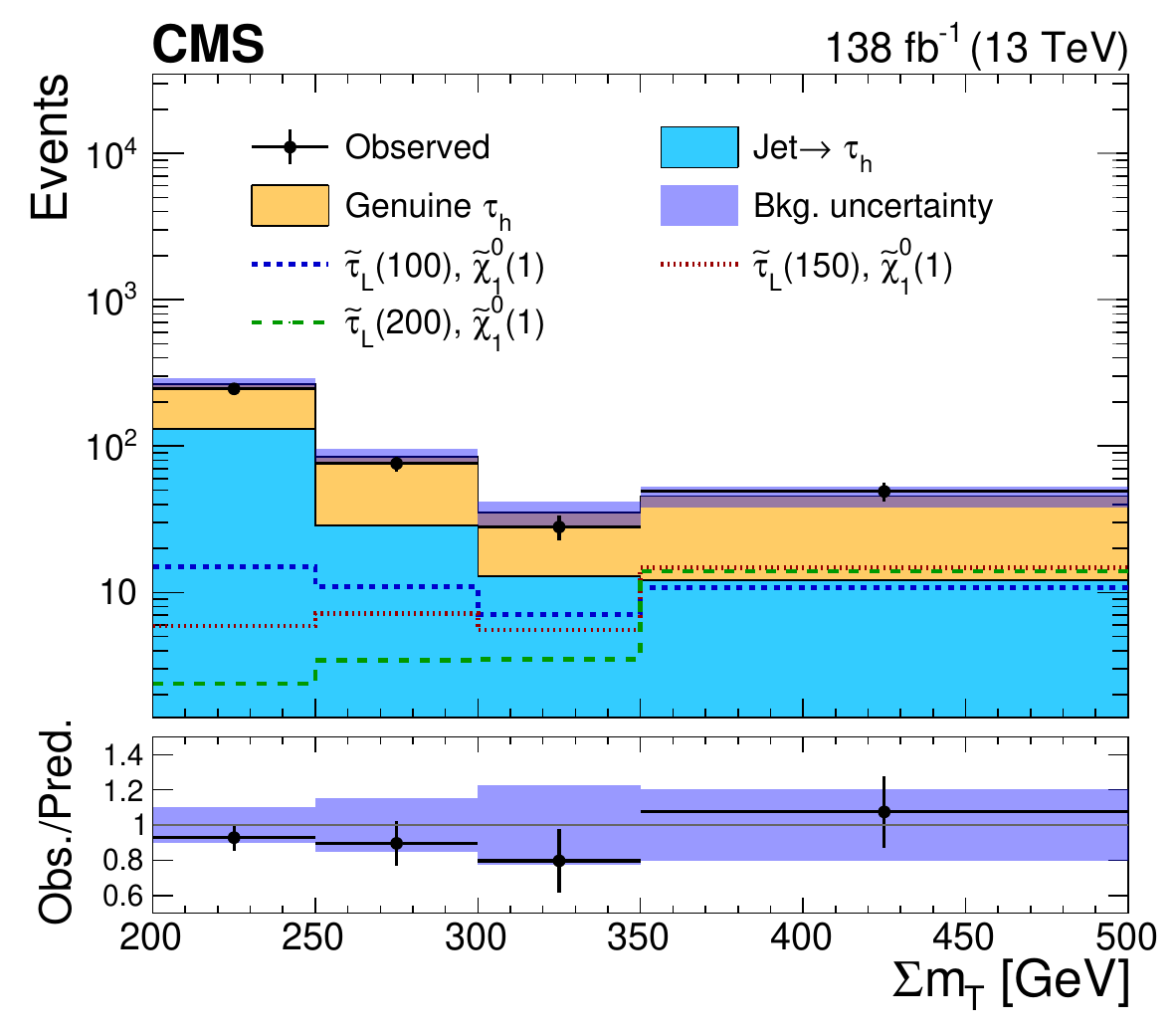} \\
\includegraphics[width=0.48\textwidth]{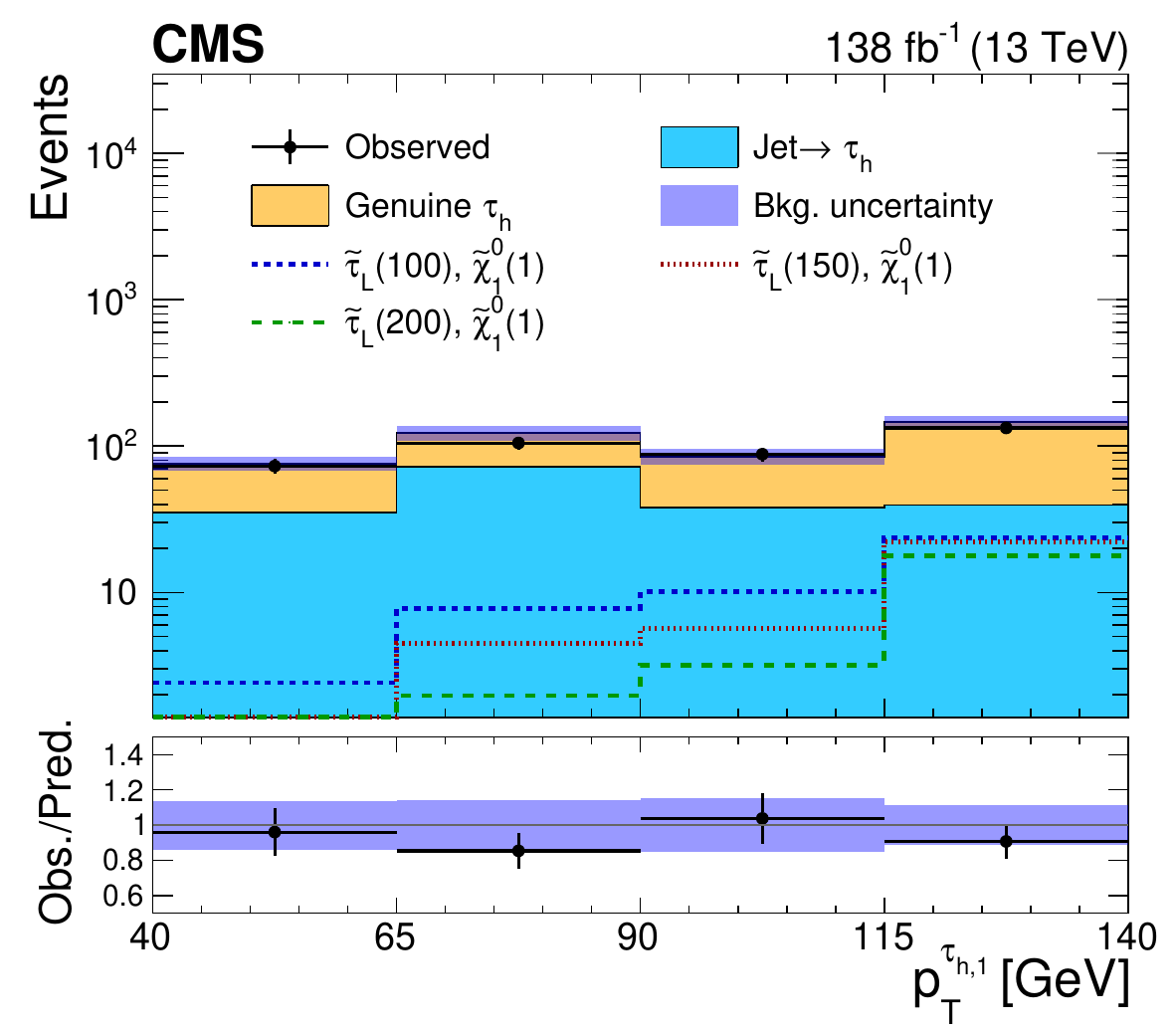}
\includegraphics[width=0.48\textwidth]{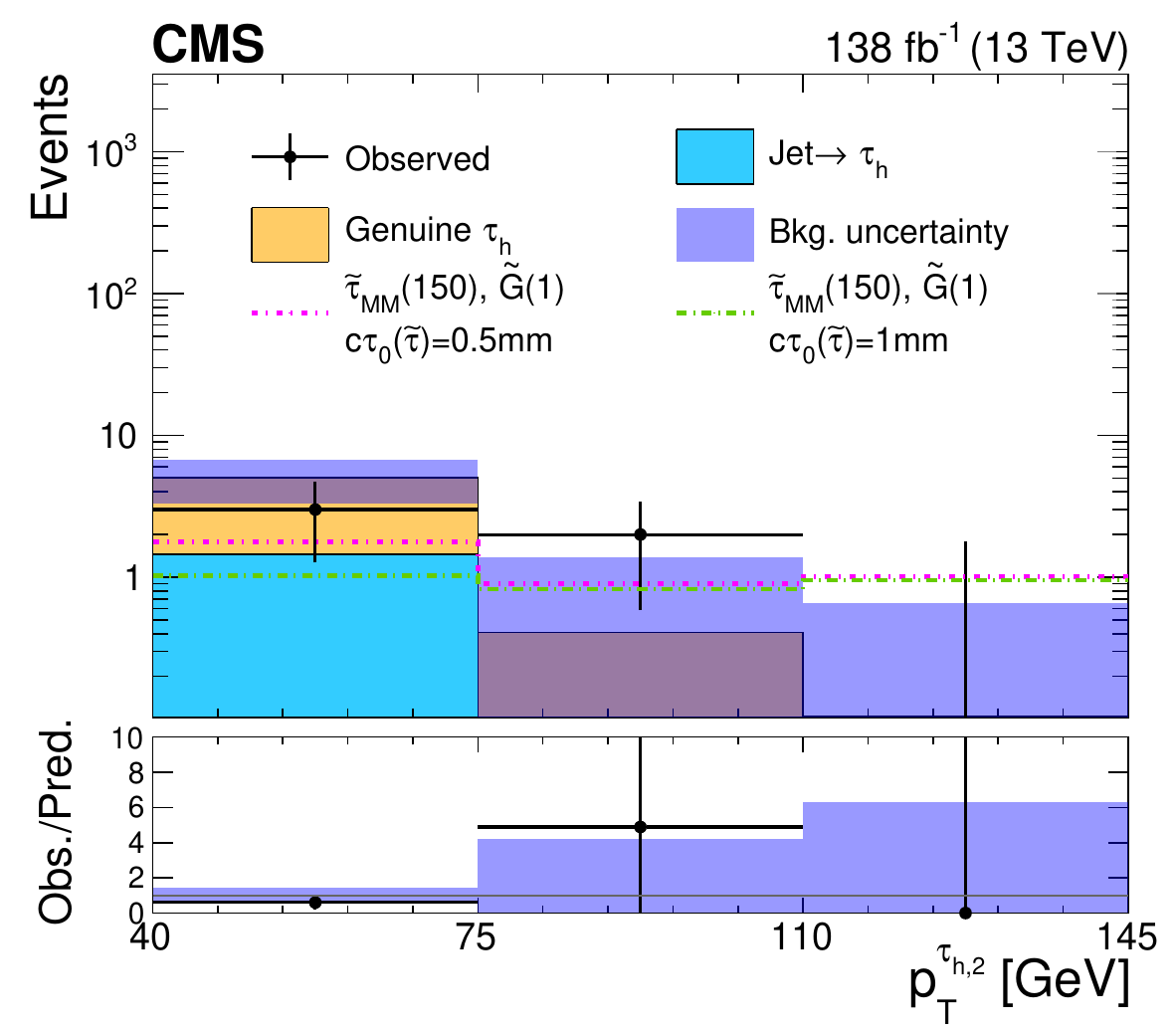}
\caption{Distributions of \mTii (upper left), \sumMT (upper right), and \taupt (lower left) for events passing the selection criteria common to all prompt SRs, and of \tautwopt (lower right) for those passing the criteria common to all displaced SRs. The last bin includes overflow events in all cases. The shaded band indicates the combined statistical and systematic uncertainty in the total SM background prediction. The lower panels show the ratio of the observed event counts to the total background prediction. Signal distributions are shown for benchmark models of \sTau pair production that are described in the text. The numbers in parentheses correspond to the masses of the \sTau and LSP in units of\GeV for the different signal models.}
\label{fig:srvariables}
\end{figure*}

The SR binning was chosen to optimize the search sensitivity, with statistical and systematic uncertainties in the signal and background predictions being taken into account in the optimization procedure. Table~\ref{tab:srdefinitions} summarizes the \sumMT, \mTii, and \taupt criteria used to define the prompt SRs, and the \tautwopt criteria used to define the displaced SRs. Figure~\ref{fig:srvariables} shows distributions of \sumMT, \mTii, and \taupt for events satisfying the selection criteria that are common to all prompt SRs in data, along with the corresponding background estimates that are obtained using the methods described in Section~\ref{sec:bkgest}. Predicted signal distributions are shown for three benchmark models of purely left-handed \sTau pair production with prompt decays with \staul masses of 100, 150, and 200\GeV and an LSP mass of 1\GeV. The distribution of \tautwopt is also shown for events satisfying the common selection criteria of the displaced SRs for data, the predicted background, and two benchmark models of long-lived \sTau pair production in the maximally-mixed scenario ($\sTau_{\text{MM}}$) with a $\sTau_{\text{MM}}$ mass of 150\GeV, an LSP mass of 1\GeV, and \ctau values of 0.5 and 1\unit{mm}.

\begin{table}[!ht]
\topcaption{\label{tab:srdefinitions} Ranges of \sumMT, \mTii, and \taupt used to define the prompt SRs for the $\nj=0$ and $\nj \geq 1$ event categories, and ranges of \tautwopt used to define the displaced SRs.}
\centering
\begin{scotch}{cccc}
\multicolumn{4}{c}{Prompt SRs} \\
\hline
SR bin & \sumMT [\GeVns{}] & \mTii [\GeVns{}] & \taupt [\GeVns{}] \\
\hline
\multicolumn{4}{c}{$\nj=0$} \\
\hline
1 & $200-250$ & $25-50$ & $<$90 \\
2 & $200-250$ & $25-50$ & $>$90 \\
3 & $200-250$ & $50-75$ & $<$90 \\
4 & $200-250$ & $50-75$ & $>$90 \\
5 & $200-250$ & $>$75  & \NA\\
6 & $250-300$ & $25-50$ &$<$90 \\
7 & $250-300$ & $25-50$ &$>$90 \\
8 & $250-300$ & $50-75$ &$<$90 \\
9 & $250-300$ & $50-75$ &$>$90 \\
10 & $250-300$ & $>$75  & \NA\\
11 & $300-350$ & $25-50$  & \NA\\
12 & $300-350$ & $50-75$  & \NA\\
13 & $300-350$ & $75-100$ & \NA\\
14 & $300-350$ & $>$100 & \NA\\
15 &    $>$350 & $25-50$  & \NA\\
16 &    $>$350 & $50-75$  & \NA\\
17 &    $>$350 & $75-100$ & \NA\\
18 &    $>$350 & $>$100 & \NA\\
\hline
\multicolumn{4}{c}{$\nj \geq 1$} \\
\hline
19 & $200-250$ & $25-50$  & \NA\\
20 & $200-250$ & $>$50  & \NA\\
21 & $250-300$ & $25-50$  & \NA\\
22 & $250-300$ & $50-75$  & \NA\\
23 & $250-300$ & $>$75  & \NA\\
24 & $300-350$ & $25-50$  & \NA\\
25 & $300-350$ & $50-75$  & \NA\\
26 & $300-350$ & $>$75  & \NA\\
27 &    $>$350 & $25-75$  & \NA\\
28 &    $>$350 & $75-100$ & \NA\\
29 &    $>$350 & $>$100 & \NA\\
\hline
\multicolumn{4}{c}{Displaced SRs} \\
\hline
SR bin & \multicolumn{3}{c}{\tautwopt [\GeVns{}]} \\
\hline
30 & \multicolumn{3}{c}{$<$110} \\
31 & \multicolumn{3}{c}{$>$110} \\
\end{scotch}
\end{table}

\section {Background estimation}
\label{sec:bkgest}

Significant contributions to the SM background in this search originate from \dytautau, \wjets, \ttbar, and diboson processes, as well as from events exclusively comprising jets produced through the strong interaction of quantum chromodynamics (QCD), which we refer to as QCD multijet events. Smaller contributions arise from single top quark production and rare SM processes, such as triboson and Higgs boson production, and top quark pair production in association with vector bosons. We rely on a method based on control samples in data to estimate the contributions of backgrounds in which one or both \tauh candidates are misidentified jets. Contributions from backgrounds with electrons or muons misidentified as \tauh candidates are negligible. We use a method known as ``embedding" for modeling backgrounds with two genuine \tauh decays~\cite{Sirunyan:2019drn}, which has the advantage that many event quantities are described by data. The background estimation methods described below are validated in dedicated background-enriched data samples that are similar in many respects but disjoint to the SRs. Separate samples are used to validate the modeling of backgrounds with prompt and displaced \tauh candidates.

\subsection{Estimation of background from misidentified jets}
\label{sec:fakeestimation}

Events with misidentified \tauh candidates, originating predominantly from QCD multijet and \wjets production, constitute the dominant background after the requirement of two \tauh candidates with high \pt. We estimate this background by extrapolating the event count in a control data sample into the SR, following the same approach as that described in~\cite{CMS:2019eln}. The control sample is selected with a relaxed \tauh identification requirement, namely the Loose working point of the \textsc{DeepTau} discriminant. We measure the fraction of misidentified \tauh candidates selected with the Loose working point that also satisfy the VVTight requirement in a QCD multijet-enriched sample of same-charge \tautau events. The fraction is found to be $\approx$10--15\%, and depends on the \pt and decay mode of the \tauh candidate, as well as on the additional activity in an event from the presence of pileup. We parameterize the measurement by the \pt and decay mode of the \tauh candidate as well as the number of reconstructed primary vertices to take these effects into account. The misidentification rate also depends on the jet flavor, \ie, whether the misidentified jet originates from the hadronization of light- or heavy-flavor quarks, or gluons, which cannot be reliably determined in data. We assign a systematic uncertainty of 30\% based on studies of the variation of the misidentification rate with jet flavor, performed with simulated samples, to account for the flavor dependence of the misidentification rate. The contribution of genuine \tauh candidates in the sideband regions selected with the relaxed identification requirement is taken into account when determining the background prediction, as described in~\cite{CMS:2019eln}.

\subsection{Estimation of backgrounds with two genuine \texorpdfstring{\tauh}{tau} decays}
\label{sec:dyestimation}

The background contribution with two genuine \tauh decays originates mainly from the \dytautau process, with smaller contributions originating from diboson production or processes with top quarks. We estimate this background using an embedding method~\cite{Sirunyan:2019drn} in which the reconstructed muons in a selected data sample of dimuon events are replaced by simulated \PGt leptons with the same kinematic properties as the original muons. This generates a set of hybrid events that rely on simulation only for the \PGt lepton decays. Consequently, they provide a better description of the underlying event, pileup, additional jets, detector noise and resolution effects compared to pure simulation samples. Since the embedded samples rely on the data for the description of these effects, no additional corrections for the pileup profile, jet energy scale, or {\cPqb} tagging efficiency are needed. Correction factors are applied to account for the efficiencies of the dimuon triggers and muon identification and isolation criteria used to select events. As the \PGt lepton decays are simulated, we apply scale factors to match the \tauh identification efficiency and energy scale in data. Correction factors are also applied to match the efficiencies of triggers used to select events for the search in data. Since the tracking efficiency in embedded events is higher than in data, scale factors are applied to account for this discrepancy. In order to save processing time, the sample selected for the detector simulation of \PGt lepton decays is restricted to events that will subsequently satisfy the selection criteria applied in analyses. This is done by applying a kinematic filter on the \pt and $\abs{\eta}$ of the visible \PGt lepton decay products. Event weights are applied to account for the bias arising from the filter requirements. These corrections are described in more detail in~\cite{Sirunyan:2019drn}.

We use a set of embedded samples in which both \PGt leptons are required to decay hadronically. We use an opposite-charge di-\tauh region in data to derive residual scale factors for the normalization of the embedded samples after all other correction factors are applied. These scale factors are designed to measure any remaining differences in the \tauh identification and trigger efficiencies between data and the embedded samples in a region that is kinematically similar to the SRs. This region consists of events passing the baseline selection, with the following additional requirements imposed on the mass and \pt of the di-\tauh system to improve the purity of genuine \tauh decays and enhance the contribution from \dytautau: $50<m_{\tautau}<90\GeV$ and $\pt^{\tautau}>50\GeV$. In order to ensure there is no overlap with the SRs and to suppress signal contamination, we require that events in this region have $\mTii<25\GeV$ or $\sumMT<200\GeV$. After subtracting the estimated contributions from misidentified \tauh events in this sample, we measure scale factors of $1.24\pm0.03$, $1.21\pm0.03$, and $1.16\pm0.02$ for 2016, 2017, and 2018 data, respectively, for the embedded events, with the uncertainties listed being statistical. We apply these scale factors, along with a conservative uncertainty corresponding to the full size of their deviations from unity, to the normalization of the embedded sample.

The genuine \tauh background prediction from the embedded sample accounts for SM events originating from processes in which the branching fractions for \PGt{}\PGt and dimuon decays are identical, \ie, \dytautau, \ttbar (with or without extra vector bosons), single top quark, and diboson processes. Small contributions from top quark events in which the {\PW} boson from the top quark decay does not decay directly into a muon and a neutrino, \eg, from $\PW\to\PGt\Pgngt$, may be included in the sample selected to undergo the embedding process, resulting in a possible overestimation of the fraction of top quark events. In order to account for this effect, we check the normalization of top quark events estimated from the embedded sample in a control region selected by requiring at least one \PQb-tagged jet, $m_{\tautau}>100\GeV$, and $\ptmiss>50\GeV$ to enrich the proportion of top quark events. Based on the level of agreement observed between data and the prediction from the embedded sample in this region, we assign an uncertainty of 10\% in the expected fraction of top quark events in the embedded sample.

The embedded sample does not account for contributions from SM Higgs boson (\PH) events, for which the \PGt{}\PGt and dimuon branching fractions are very different. We therefore include the estimated contribution from SM $\PH\to\PGt\PGt$ events from simulation in the total estimate of the genuine \tauh background. We find that the background contribution from SM $\PH\to\PGt\PGt$ events is small compared to the other backgrounds.

\section{Systematic uncertainties}
\label{sec:sysunc}

\begin{table*}[htbp]
\topcaption{\label{tab:syst}Uncertainties in the analysis affecting signal and the SM backgrounds. The numbers indicate the percentage effect of propagating $\pm1$-standard deviation variations of the respective sources of uncertainty on the predicted signal and background yields, prior to a fit to the data. The ranges shown for signal refer to a representative benchmark model of \staul pair production with $m(\staul)=150\GeV$, $m(\lsp)=1\GeV$.}
\centering
\begin{scotch}{lccc}
 & \multicolumn{3}{c}{Uncertainty [\%]} \\
\cline{2-4}
Source & Genuine \tauh& Misidentified \tauh& Signal\\
\hline
Statistical  & 8.3--141 & 5.0--100 & 6.3--52 \\
\tauh ID efficiency  & 7.2--7.8 & \NA & 6.2--6.4 \\
\tauh ID vs. displacement  & \NA & \NA & 3.0 \\
\tauh trigger efficiency  & 3.1--4.2 & \NA & 6.9--14 \\
\tauh energy scale  & 0.1--35 & \NA & 1.6--44 \\
\tauh misidentification rate  & \NA & 30--56 & \NA \\
\met trigger efficiency  & 1.0 & \NA & 1.5 \\
Embedded normalization  & 19 & \NA & \NA \\
Embedded top quark fraction  & 1.0--3.8 & \NA & \NA \\
Jet energy scale  & \NA & \NA & 0.7--32 \\
Jet energy resolution  & \NA & \NA & 1.3--55 \\
Unclustered energy  & \NA & \NA & 0.5--32 \\
{\cPqb} tagging  & \NA & \NA & 0.2--1.1 \\
Pileup  & \NA & \NA & 1.0--28 \\
Integrated luminosity  & \NA & \NA & 1.6 \\
ISR  & \NA & \NA & 0.1--16 \\
Renormalization/factorization scales  & \NA & \NA & 0.4--3.6 \\
\end{scotch}
\end{table*}

The dominant uncertainties in the background estimates are the statistical uncertainty driven by the limited event counts in the data sidebands or embedded samples used to obtain the estimates, and the systematic uncertainty (30\%) assigned to the estimate of the \tauh misidentification rate that accounts for its dependence on jet flavor.

Because we use embedded events to estimate the background with two genuine \tauh decays, the prediction is less affected by systematic uncertainties than in the case of wholly simulated samples. For this background, we propagate uncertainties related to the trigger efficiency, \tauh identification efficiency, and \tauh energy scale. We also assign additional uncertainties as discussed in Section~\ref{sec:dyestimation}: a 10\% uncertainty in the expected fraction of top quark events in the embedded samples estimated from simulation, and a normalization uncertainty determined by the full size of the deviations from unity of the normalization scale factors derived from the \dytautau control region, weighted over the three years by the fractions of the integrated luminosity collected in each year. A 20\% normalization uncertainty is assigned to the small contribution from SM $\PH\to\PGt\PGt$ events.

For the signal prediction obtained from simulation, we propagate uncertainties in the trigger efficiency, \tauh identification efficiency, \tauh energy scale, {\cPqb} tagging efficiency, pileup reweighting, jet energy scale and resolution, and unclustered energy. We also take into account the uncertainty in the integrated luminosity measurement. The integrated luminosities for the 2016, 2017, and 2018 data-taking years have 1.2--2.5\% individual uncertainties~\cite{CMS-LUM-17-003,CMS-PAS-LUM-17-004,CMS-PAS-LUM-18-002}, while the overall uncertainty for the 2016--2018 period is 1.6\%. Uncertainties related to the renormalization and factorization scales, and to the modeling of ISR, are propagated to the signal prediction as well. Since the \tauh identification and trigger efficiency correction factors applied to simulation, which are obtained from samples of \dytautau events with promptly produced \PGt leptons, do not account for the dependence of the \tauh candidate selection on the displacement of the decay position, we assign an additional uncertainty for this effect in signal events. The uncertainty is assessed via a comparison of \tauh impact parameter distributions between data and simulation in a control region that is mainly populated by \dytautau events, and the expected displacement of signal events for different \sTau lifetimes. In order to probe the tails of the \tauh impact parameter distributions, which we expect to be populated by signal events with significant displacement, the data-to-simulation ratios observed in the control region at lower values of \dxy significance and IP3D are extrapolated to higher values via a linear fit when deriving this uncertainty. The size of the uncertainty ranges from 3\% for promptly decaying \sTau, to 45\% for \sTau with $\ctau=2.5\unit{mm}$.

We treat statistical uncertainties as uncorrelated, while systematic uncertainties related to the same modeling effect are taken to be correlated across processes and data-taking periods. Table~\ref{tab:syst} lists the ranges of the uncertainty in the predicted yields for background and a benchmark signal model of \staul pair production with $m(\staul)=150\GeV$, $m(\lsp)=1\GeV$ across all SRs corresponding to different sources.

\section{Results and interpretation}
\label{sec:results}

\begin{figure*}[hp!]
\centering
\includegraphics[width=0.9\textwidth]{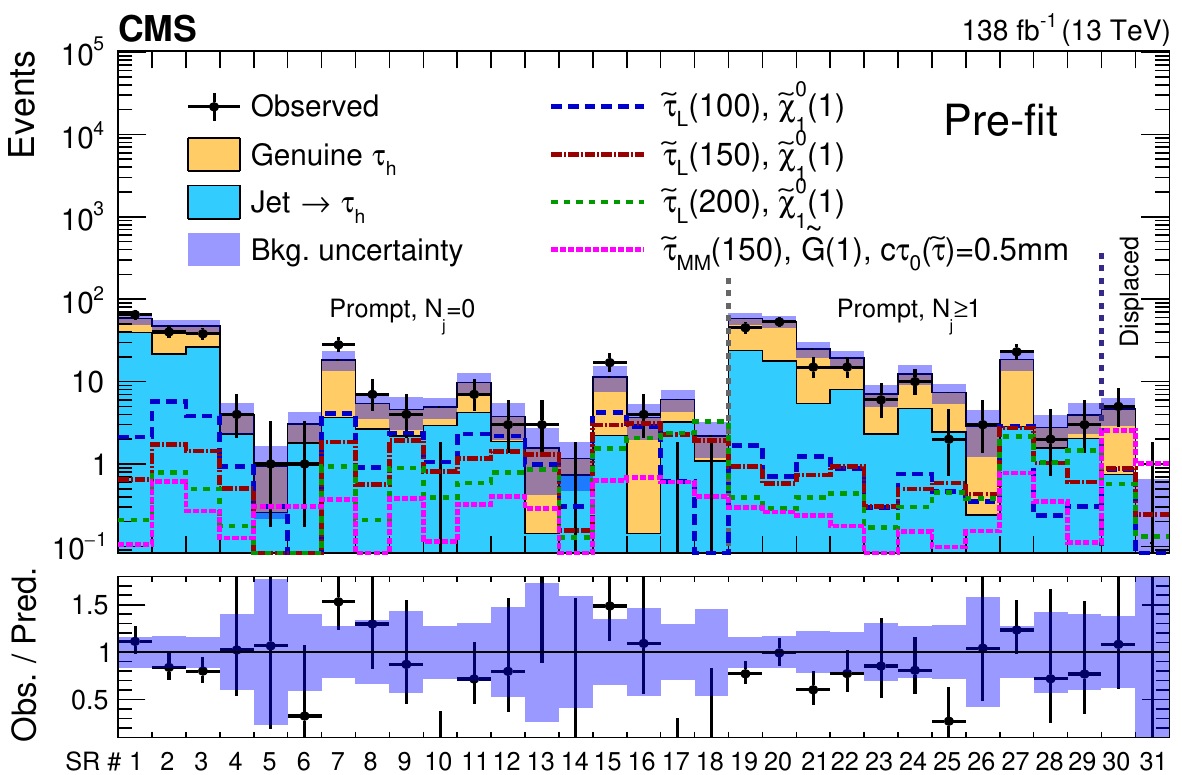} \\
\includegraphics[width=0.9\textwidth]{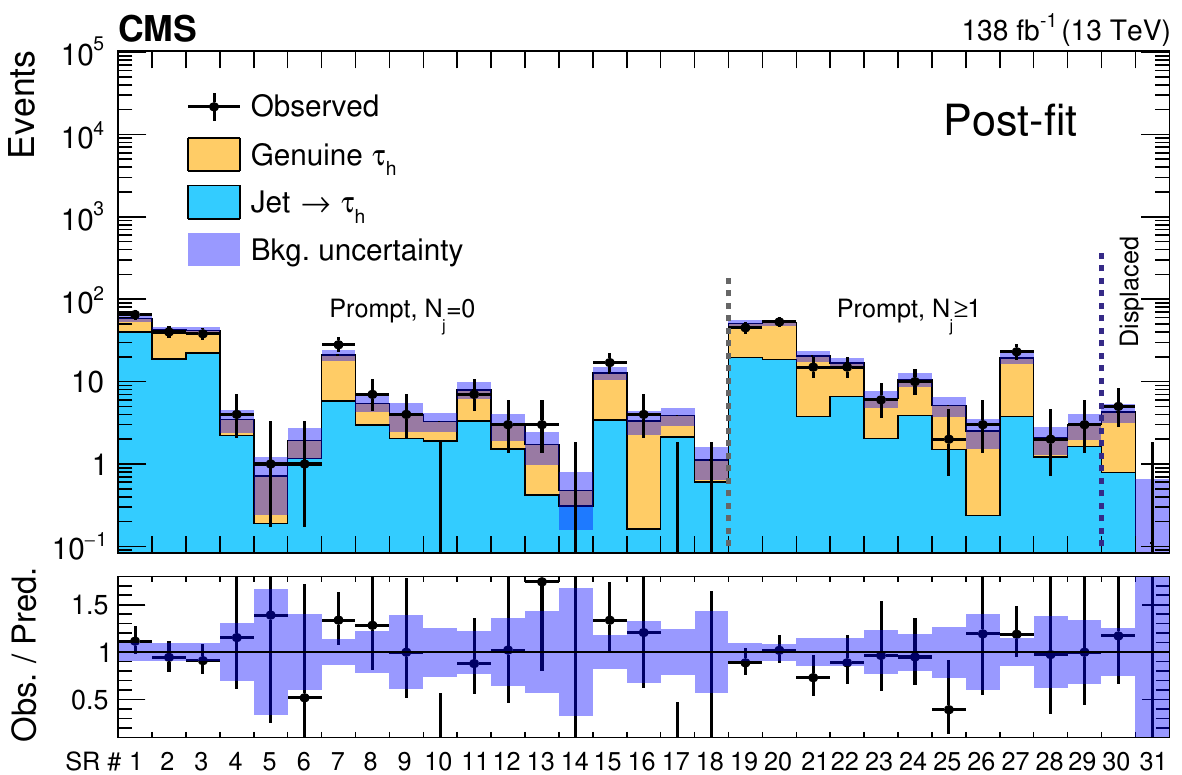}
\caption{\label{fig:sryields} Event counts and predicted yields in each SR for the SM background before (upper) and after (lower) a maximum likelihood fit to the data under the background-only hypothesis. The yields expected for 3 benchmark models of left-handed \sTau pair production assuming prompt \sTau decays, and one model of long-lived \sTau pair production in the maximally mixed scenario ($\sTau_{\text{MM}}$) are overlaid in the pre-fit case. The numbers in parentheses correspond to the masses of the \sTau and LSP in units of\GeV for the different signal models. The lower panels show the ratio of the observed event counts to the total SM background prediction. The first 29 bins correspond to the prompt SRs, while bins 30 and 31 correspond to the displaced SRs, as labeled in Table~\ref{tab:srdefinitions}.}
\end{figure*}

Observed and predicted event yields for each SR are shown in Fig.~\ref{fig:sryields} (upper) and summarized in Table~\ref{tab:results} for the combined 2016--2018 data set. Figure~\ref{fig:sryields} (lower) shows the background predictions after a maximum likelihood fit to the data under the background-only hypothesis. The likelihood function is constructed from the observed and predicted event yields in all 31 SRs, with the uncertainties in the predicted yields incorporated as nuisance parameters in the fit. The normalization uncertainties affecting background and signal predictions are generally assumed to be log-normally distributed. For statistical uncertainties limited by small event counts in the embedded or simulation samples, or in the sideband regions in the data used to estimate the misidentified \tauh background, we use a Poisson distribution. The nuisance parameters are allowed to vary within their uncertainties in the fit. The uncertainties in the background predictions are reduced after the fit to the data. In general, the data are consistent with the prediction for the SM background. The most significant discrepancies between the data and the predicted background occur in SR bins 10 and 17, where we observe a deficit at approximately the 2-standard deviation level with respect to the pre-fit predictions, while a smaller deficit, at about the 1.2-standard deviation level, is observed in SR bin 18. Signal predictions are also shown in Fig.~\ref{fig:sryields} (upper) for three benchmark models of left-handed prompt \sTau pair production with \sTau masses of 100, 150, and 200\GeV for an LSP mass of 1\GeV, and for one of long-lived \sTau pair production in the maximally mixed scenario with a \sTau mass of 150\GeV, an LSP mass of 1\GeV, and $c\tau_{0}(\sTau)=0.5$\unit{mm}.

\begin{table*}[hp!]
\topcaption{\label{tab:results} Predicted SM background yields, observed event counts, and predicted signal yields for two benchmark models with a \sTau mass of 150\GeV and an LSP mass of 1\GeV, in all prompt and displaced SRs as labeled in Table~\ref{tab:srdefinitions}. For the prompt signal model shown, we assume left-handed \sTau pair production, while for the displaced signal model we assume a maximally mixed scenario and $c\tau_{0}(\sTau)=0.5$\unit{mm}. The uncertainties listed are the sum in quadrature of the statistical and systematic components. For any estimate with no events in the data sideband, embedded, or simulation sample corresponding to a given SR selection, we provide the one standard deviation upper bound evaluated for that estimate.}
\centering
\cmsTable{
\renewcommand{\arraystretch}{1.3}
\begin{scotch}{ccccccc}
SR bin & Genuine \tauh & Misidentified \tauh & Total SM & Observed & Prompt signal & Displaced \tauh signal \\
\hline
1  & $18.8 \pm 4.4$ & $39.6 \pm 8.5$ & $58.4 \pm 9.6$ & 65 & $0.6 \pm 0.1$ & $0.1 \pm 0.1$ \\ 
2  & $25.9 \pm 6.2$ & $21.8 \pm 5.5$ & $47.7 \pm 8.2$ & 40 & $1.7 \pm 0.3$ & $0.6 \pm 0.2$ \\ 
3  & $21.4 \pm 4.9$ & $26.1 \pm 6.0$ & $47.5 \pm 7.7$ & 38 & $1.4 \pm 0.3$ & $0.3 \pm 0.1$ \\ 
4  & $1.3^{+1.0}_{-0.6}$ & $2.1^{+1.5}_{-1.4}$ & $3.4^{+1.8}_{-1.5}$ & 4 & $0.5 \pm 0.1$ & $0.1 \pm 0.1$ \\ 
5  & $0.5^{+0.5}_{-0.3}$ & $0.1^{+0.7}_{-0.1}$ & $0.6^{+0.9}_{-0.4}$ & 1 & $0.0 \pm 0.0$ & $<$0.6 \\ 
6  & $1.0^{+0.9}_{-0.6}$ & $1.6^{+1.1}_{-0.9}$ & $2.5^{+1.5}_{-1.1}$ & 1 & $0.0 \pm 0.0$ & $<$0.6 \\ 
7  & $14.6 \pm 3.9$ & $3.7 \pm 3.0$ & $18.3 \pm 4.9$ & 28 & $1.8 \pm 0.3$ & $0.4 \pm 0.2$ \\ 
8  & $2.5^{+1.1}_{-0.9}$ & $2.7 \pm 1.5$ & $5.2^{+1.9}_{-1.8}$ & 7 & $0.6 \pm 0.1$ & $0.1 \pm 0.0$ \\ 
9  & $2.1^{+1.3}_{-1.0}$ & $2.2 \pm 1.6$ & $4.3^{+2.1}_{-1.9}$ & 4 & $1.9 \pm 0.3$ & $0.4 \pm 0.1$ \\ 
10  & $1.8^{+0.9}_{-0.7}$ & $2.7^{+1.3}_{-1.1}$ & $4.5^{+1.6}_{-1.3}$ & 0 & $0.8 \pm 0.2$ & $0.1 \pm 0.1$ \\ 
11  & $5.2^{+2.1}_{-1.8}$ & $4.2 \pm 2.2$ & $9.4^{+3.1}_{-2.9}$ & 7 & $1.2 \pm 0.2$ & $0.3 \pm 0.1$ \\ 
12  & $1.5^{+1.3}_{-0.9}$ & $1.7^{+1.4}_{-1.2}$ & $3.2^{+1.9}_{-1.5}$ & 3 & $1.4 \pm 0.3$ & $0.4 \pm 0.1$ \\ 
13  & $1.1^{+1.1}_{-0.6}$ & $0.1^{+1.0}_{-0.1}$ & $1.1^{+1.5}_{-0.6}$ & 3 & $1.3 \pm 0.2$ & $0.3 \pm 0.1$ \\ 
14  & $0.2^{+0.5}_{-0.2}$ & $0.5^{+0.8}_{-0.4}$ & $0.7^{+0.9}_{-0.5}$ & 0 & $0.2 \pm 0.1$ & $0.0 \pm 0.0$ \\ 
15  & $8.9^{+2.9}_{-2.6}$ & $2.2 \pm 2.6$ & $11.1^{+3.9}_{-3.7}$ & 17 & $3.0 \pm 0.4$ & $0.6 \pm 0.2$ \\ 
16  & $3.2^{+1.6}_{-1.3}$ & $<$1.0 & $3.2^{+1.9}_{-1.3}$ & 4 & $3.1 \pm 0.5$ & $0.7 \pm 0.2$ \\ 
17  & $2.5^{+1.4}_{-1.1}$ & $3.0^{+1.5}_{-1.3}$ & $5.5^{+2.1}_{-1.7}$ & 0 & $2.3 \pm 0.4$ & $0.6 \pm 0.2$ \\ 
18  & $0.7^{+1.0}_{-0.5}$ & $0.9^{+0.8}_{-0.5}$ & $1.6^{+1.3}_{-0.7}$ & 0 & $1.9 \pm 0.3$ & $0.4 \pm 0.1$ \\ 
19  & $34.6 \pm 7.9$ & $23.8 \pm 5.5$ & $58.4 \pm 9.6$ & 45 & $0.9 \pm 0.2$ & $0.3 \pm 0.1$ \\ 
20  & $35.7 \pm 7.7$ & $17.7 \pm 4.8$ & $53.4 \pm 9.0$ & 53 & $0.6 \pm 0.1$ & $0.3 \pm 0.1$ \\ 
21  & $19.5 \pm 4.9$ & $5.4 \pm 2.6$ & $24.9 \pm 5.5$ & 15 & $0.7 \pm 0.2$ & $0.2 \pm 0.1$ \\ 
22  & $11.4 \pm 3.0$ & $8.0 \pm 3.0$ & $19.4 \pm 4.2$ & 15 & $0.9 \pm 0.2$ & $0.2 \pm 0.1$ \\ 
23  & $4.5^{+1.5}_{-1.3}$ & $2.3 \pm 1.6$ & $6.8^{+2.2}_{-2.1}$ & 6 & $0.3 \pm 0.1$ & $0.0 \pm 0.0$ \\ 
24  & $7.3^{+2.7}_{-2.4}$ & $4.7 \pm 2.4$ & $12.0^{+3.6}_{-3.4}$ & 10 & $0.5 \pm 0.1$ & $0.2 \pm 0.1$ \\ 
25  & $4.6^{+1.9}_{-1.6}$ & $2.3^{+1.4}_{-1.2}$ & $6.9^{+2.3}_{-2.0}$ & 2 & $0.6 \pm 0.1$ & $0.1 \pm 0.1$ \\ 
26  & $2.3^{+1.3}_{-1.0}$ & $<$1.3 & $2.3^{+1.9}_{-1.0}$ & 3 & $0.4 \pm 0.1$ & $0.2 \pm 0.1$ \\ 
27  & $15.8 \pm 4.1$ & $2.8 \pm 3.1$ & $18.6 \pm 5.2$ & 23 & $2.8 \pm 0.4$ & $0.8 \pm 0.2$ \\ 
28  & $0.9^{+0.9}_{-0.5}$ & $1.4^{+1.1}_{-0.9}$ & $2.3^{+1.5}_{-1.0}$ & 2 & $1.0 \pm 0.2$ & $0.4 \pm 0.1$ \\ 
29  & $1.5^{+1.2}_{-0.8}$ & $1.9^{+1.4}_{-1.1}$ & $3.4^{+1.9}_{-1.4}$ & 3 & $0.6 \pm 0.1$ & $0.1 \pm 0.1$ \\ 
30  & $3.6^{+1.5}_{-1.3}$ & $0.7^{+1.3}_{-0.7}$ & $4.3^{+2.0}_{-1.5}$ & 5 & $0.9 \pm 0.2$ & $2.6 \pm 0.7$ \\ 
31  & $<$0.5 & $<$0.4 & $0.0^{+0.7}_{-0.0}$ & 0 & $0.2 \pm 0.1$ & $1.0 \pm 0.3$ \\ 
\end{scotch}
}
\end{table*}

We use the results to set upper limits on the cross section for the production of \PSGt pairs in the context of simplified models~\cite{Simp,Alwall:2008ag,Alwall:2008va,Alves:2011wf} using all of the 31 exclusive SRs in a full statistical combination. The 95\% \CL upper limits on SUSY production cross sections are calculated using a modified frequentist approach with the \CLs criterion~\cite{Junk:1999kv,Read:2002hq,CMS-NOTE-2011-005}. An asymptotic approximation is used for the test statistic~\cite{Cowan:2010js}.

\begin{figure*}[tbp!]
\centering
\includegraphics[width=0.48\textwidth]{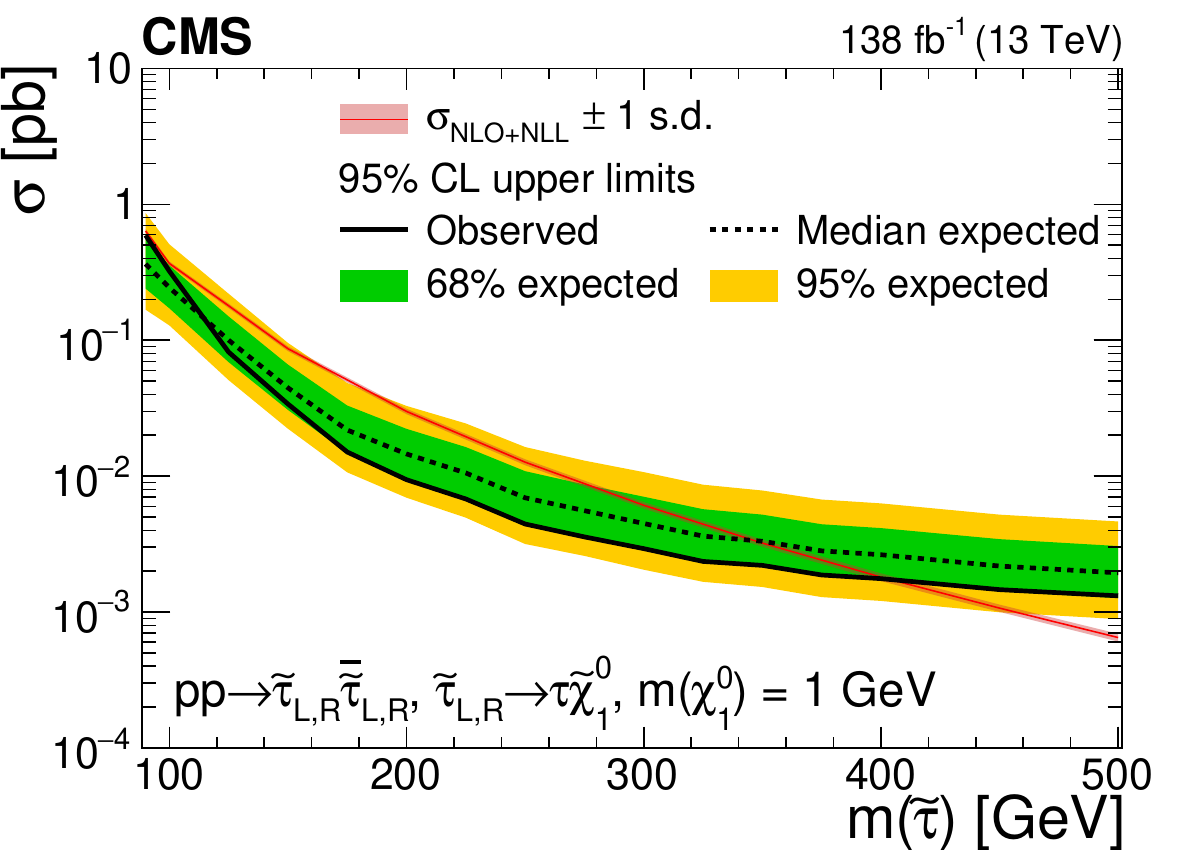}
\includegraphics[width=0.48\textwidth]{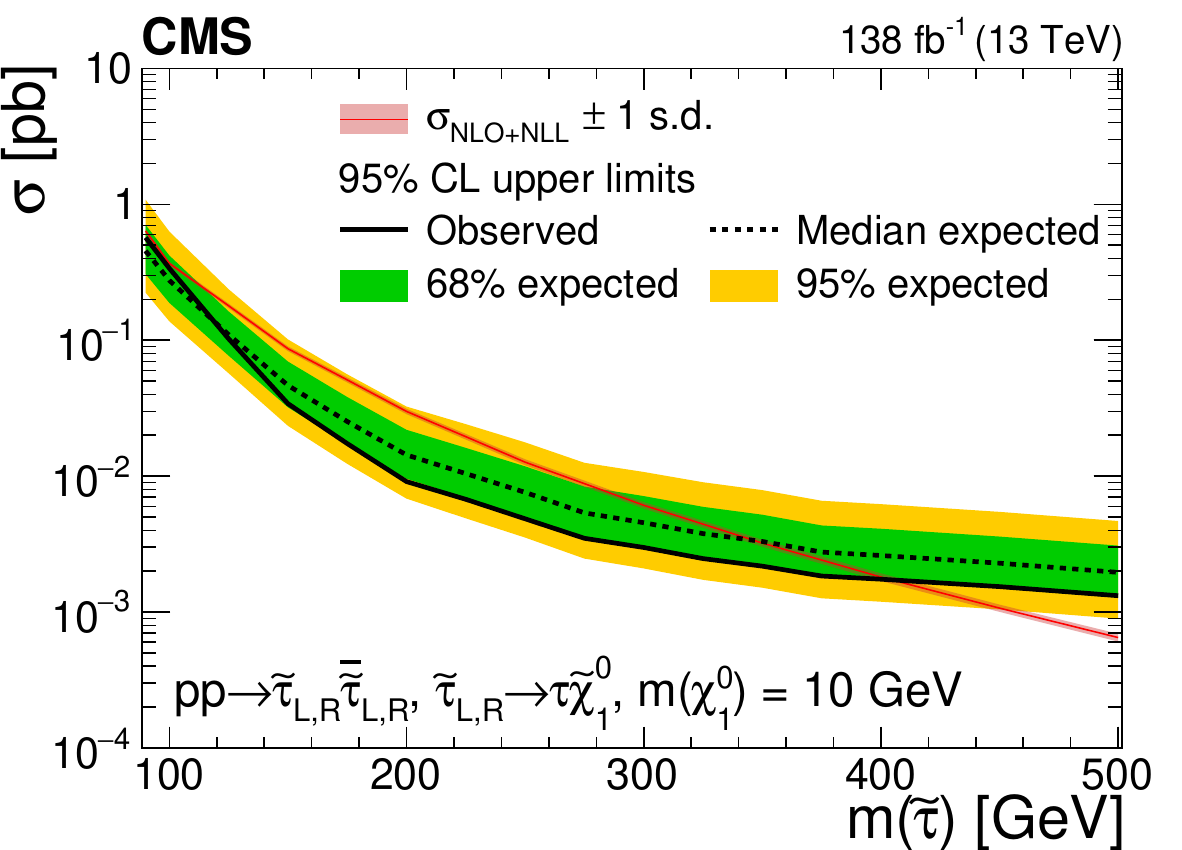} \\
\includegraphics[width=0.48\textwidth]{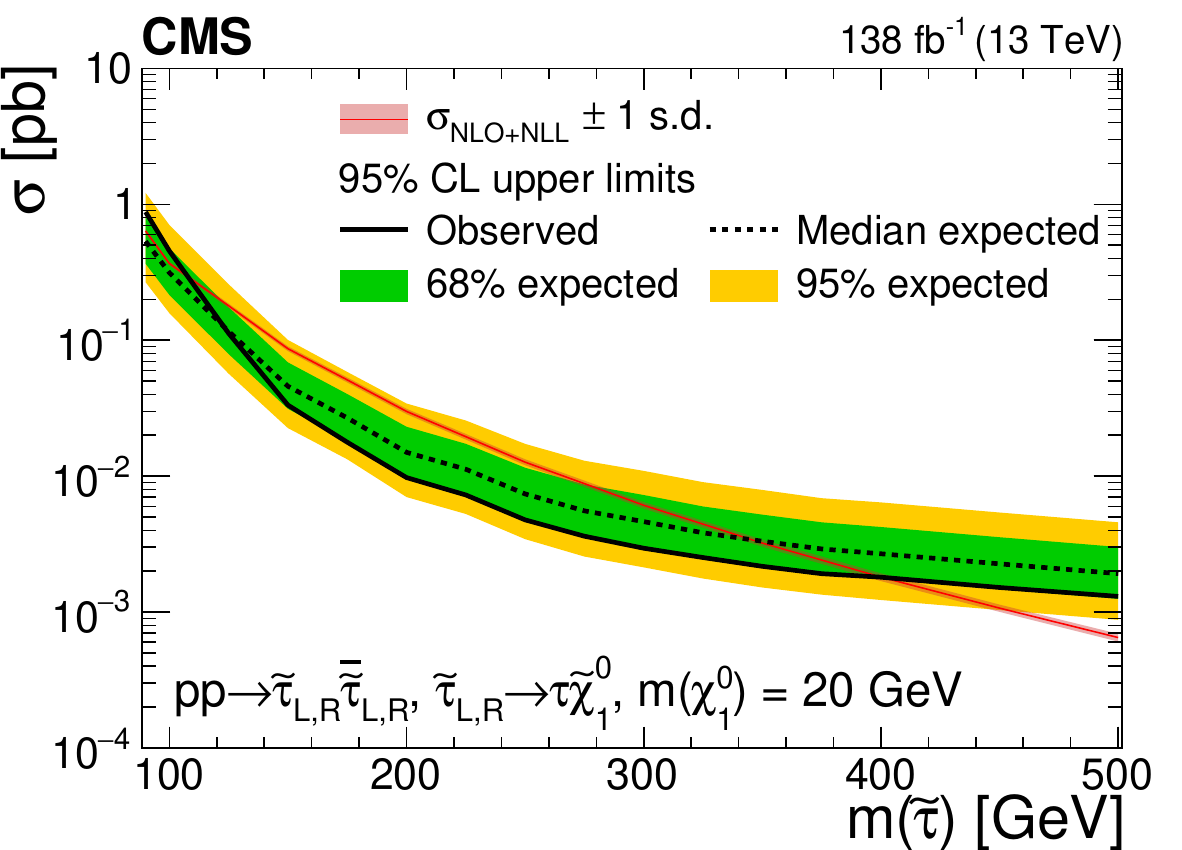}
\includegraphics[width=0.48\textwidth]{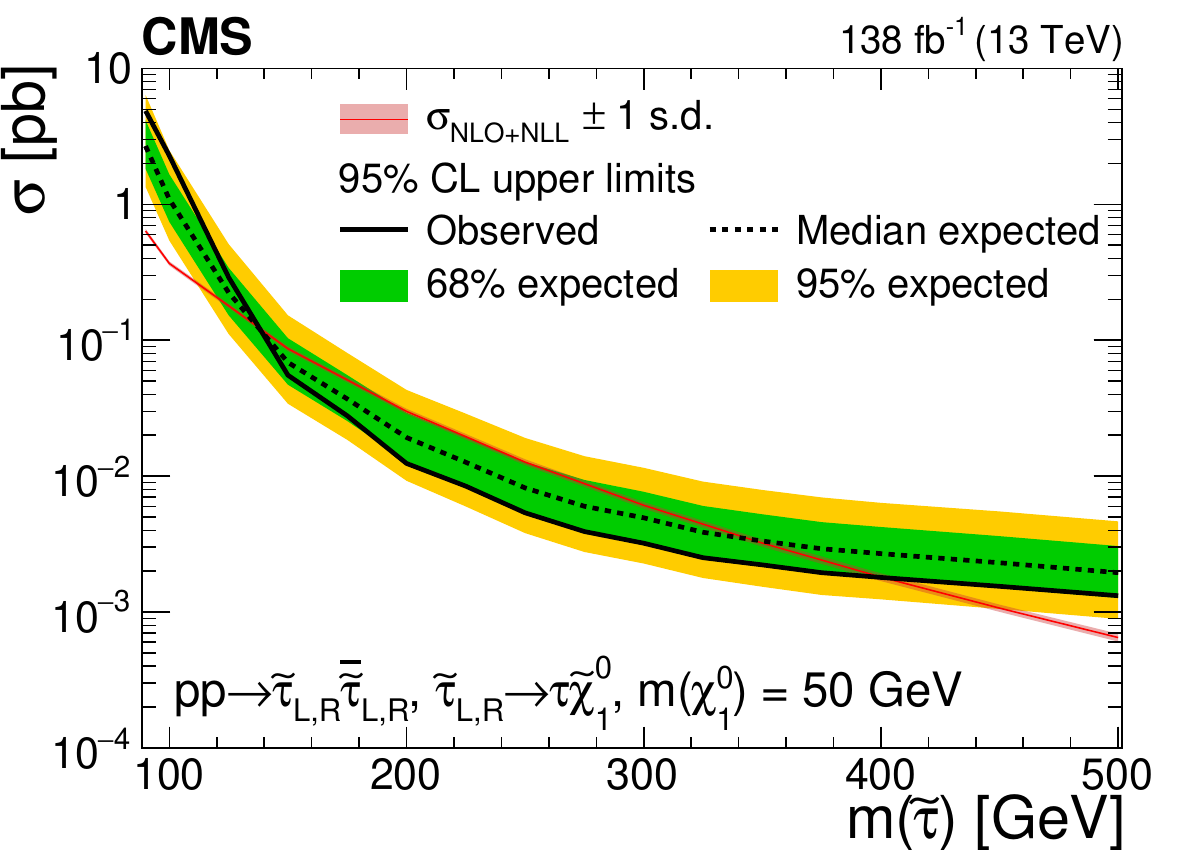}
\caption{Expected and observed 95\% \CL cross section upper limits as functions of the \sTau mass in the degenerate \sTau scenario for \lsp masses of 1, 10, 20, and 50\GeV (upper left to lower right). The inner (green) band and the outer (yellow) band indicate the regions containing 68 and 95\%, respectively, of the distribution of limits expected under the background-only hypothesis. The red line and thin shaded band indicate the NLO+NLL prediction for the signal production cross section calculated with \textsc{Resummino}~\cite{Beenakker:1999xh,Fuks:2013lya}, and its uncertainty.}
\label{fig:limitsdegen}
\end{figure*}

\begin{figure*}[htbp!]
\centering
\includegraphics[width=0.48\textwidth]{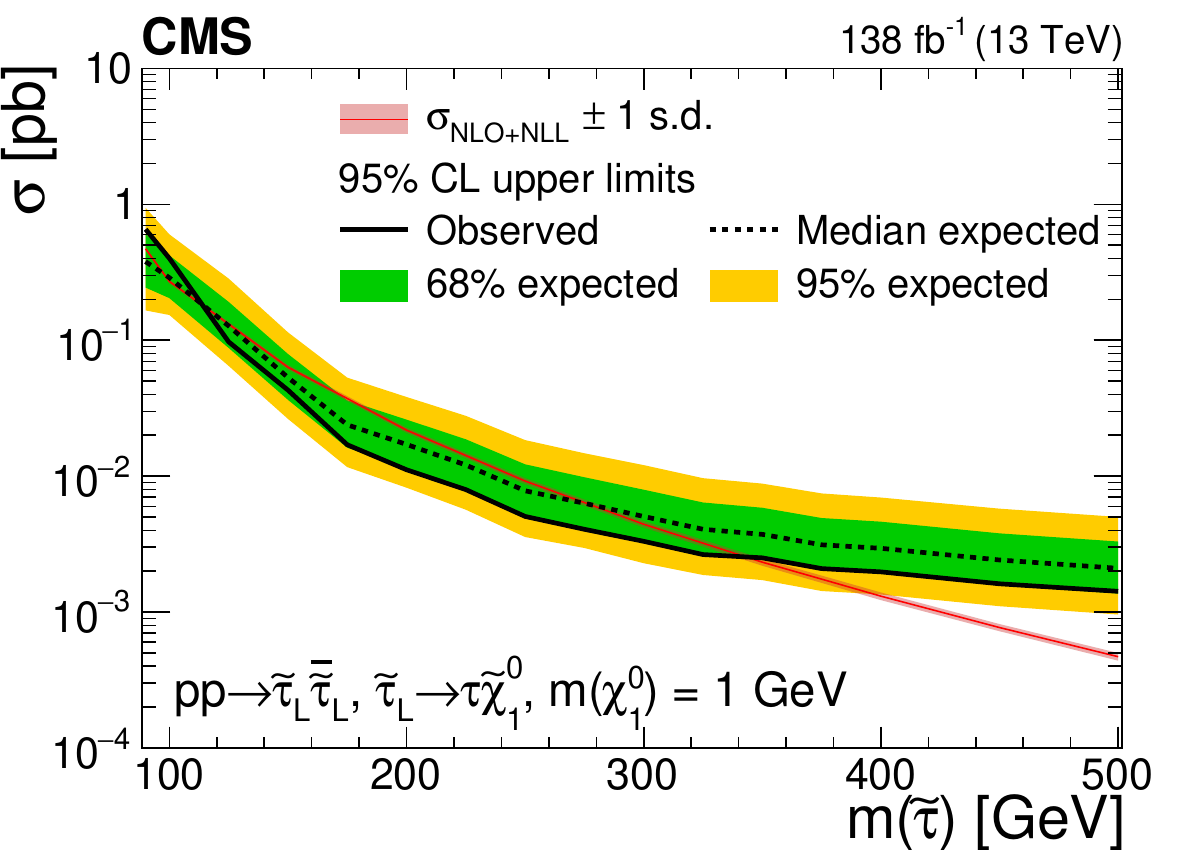}
\includegraphics[width=0.48\textwidth]{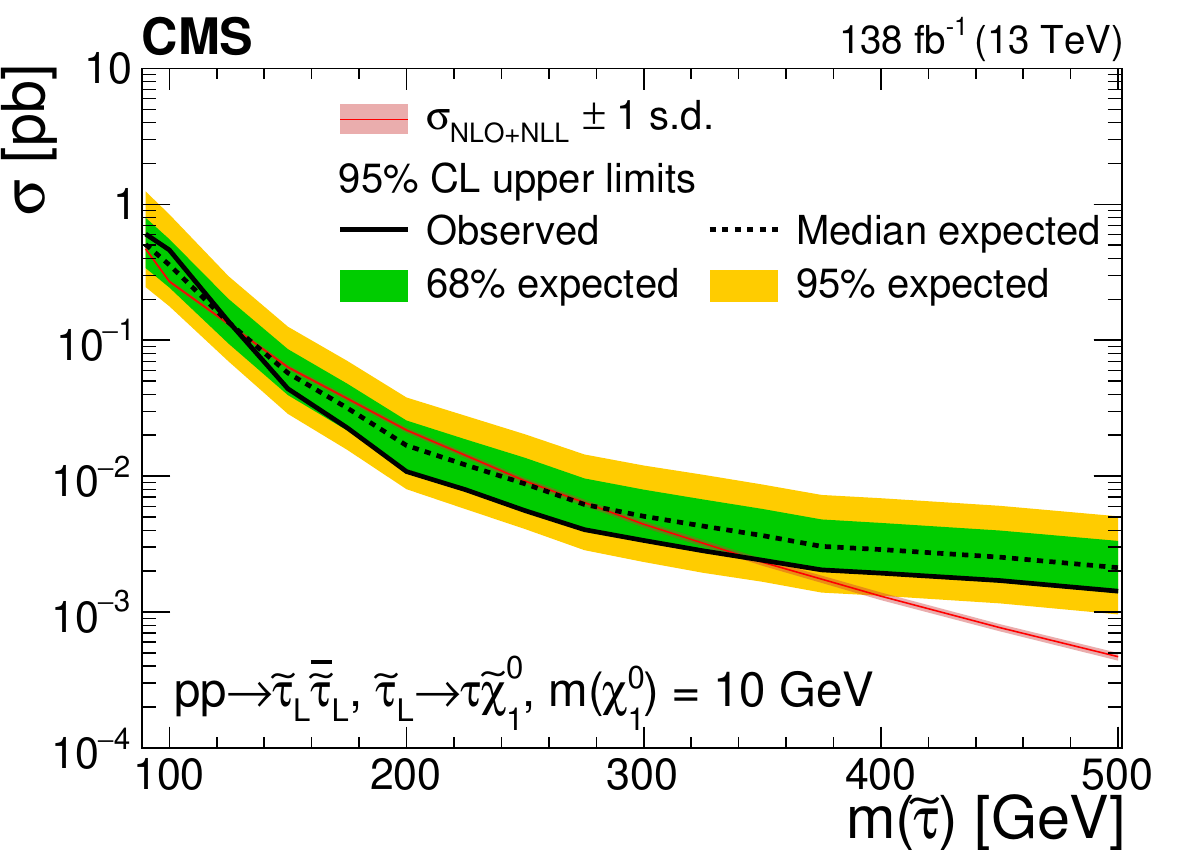} \\
\includegraphics[width=0.48\textwidth]{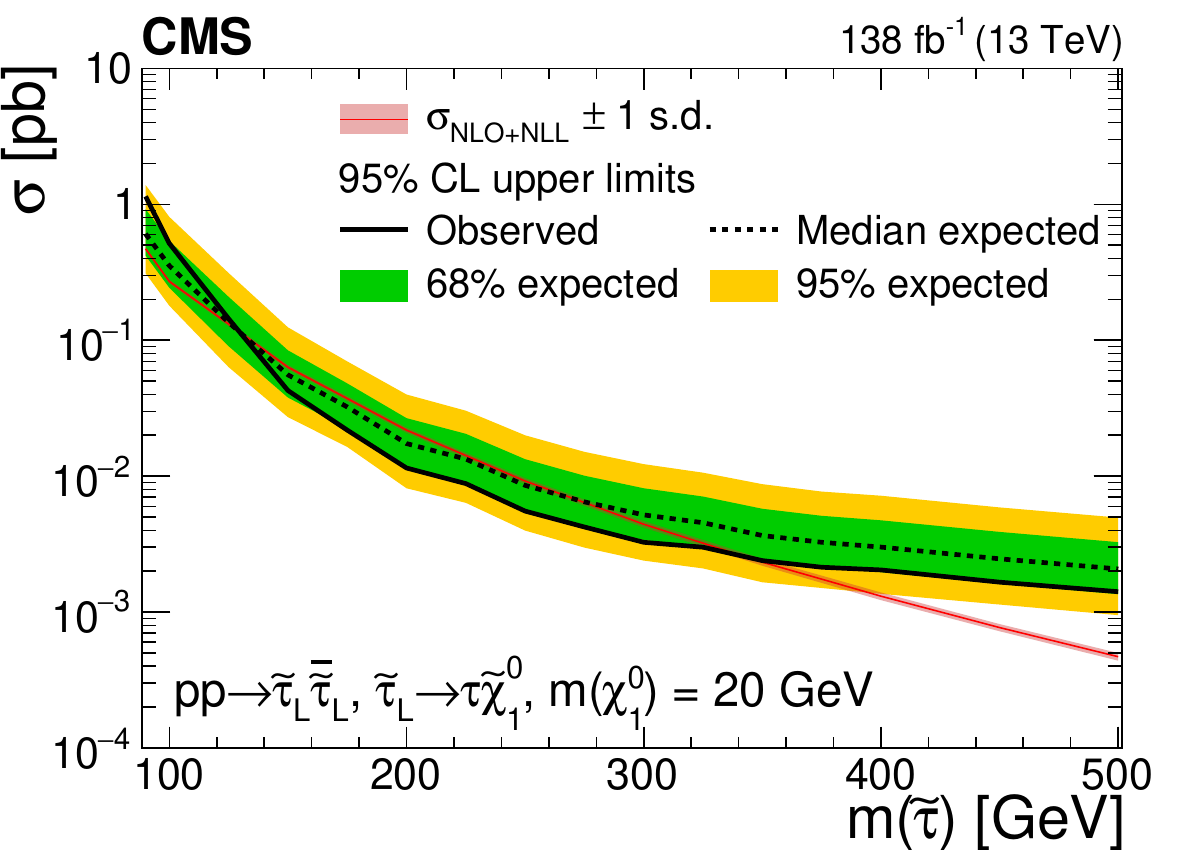}
\includegraphics[width=0.48\textwidth]{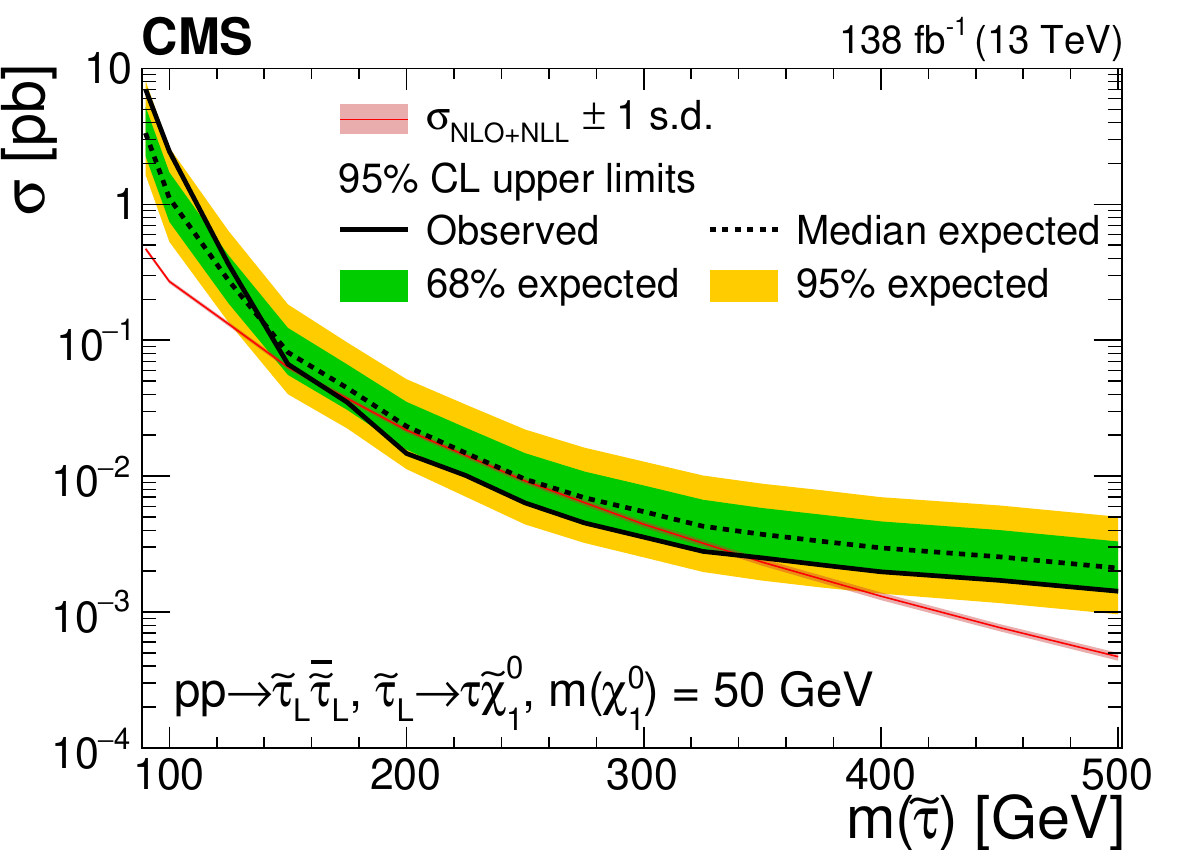}
\caption{Expected and observed 95\% \CL cross section upper limits as functions of the \sTau mass in the purely left-handed \sTau scenario for \lsp masses of 1, 10, 20, and 50\GeV (upper left to lower right). The inner (green) band and the outer (yellow) band indicate the regions containing 68 and 95\%, respectively, of the distribution of limits expected under the background-only hypothesis. The red line and thin shaded band indicate the NLO+NLL prediction for the signal production cross section calculated with \textsc{Resummino}~\cite{Fuks:2013lya}, and its uncertainty.}
\label{fig:limitsleft}
\end{figure*}

\begin{figure*}[htbp!]
\centering
\includegraphics[width=0.48\textwidth]{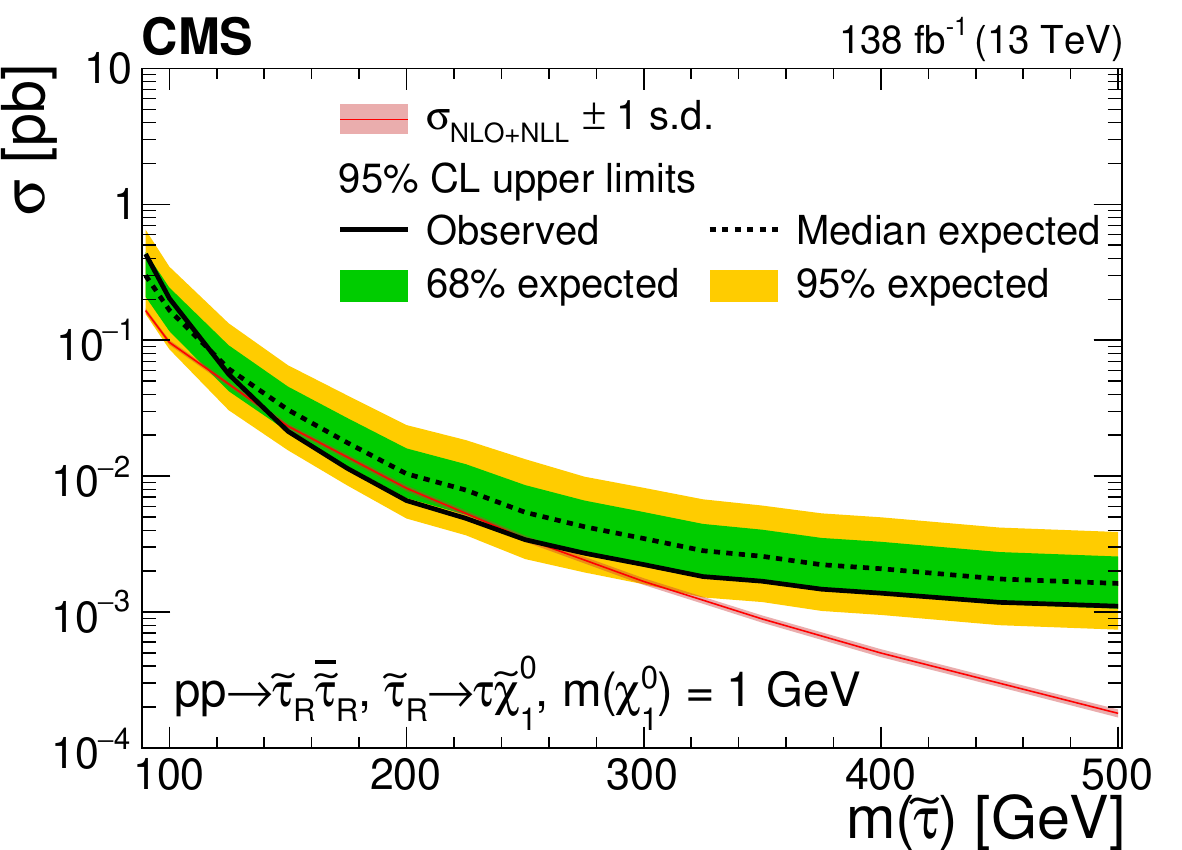}
\includegraphics[width=0.48\textwidth]{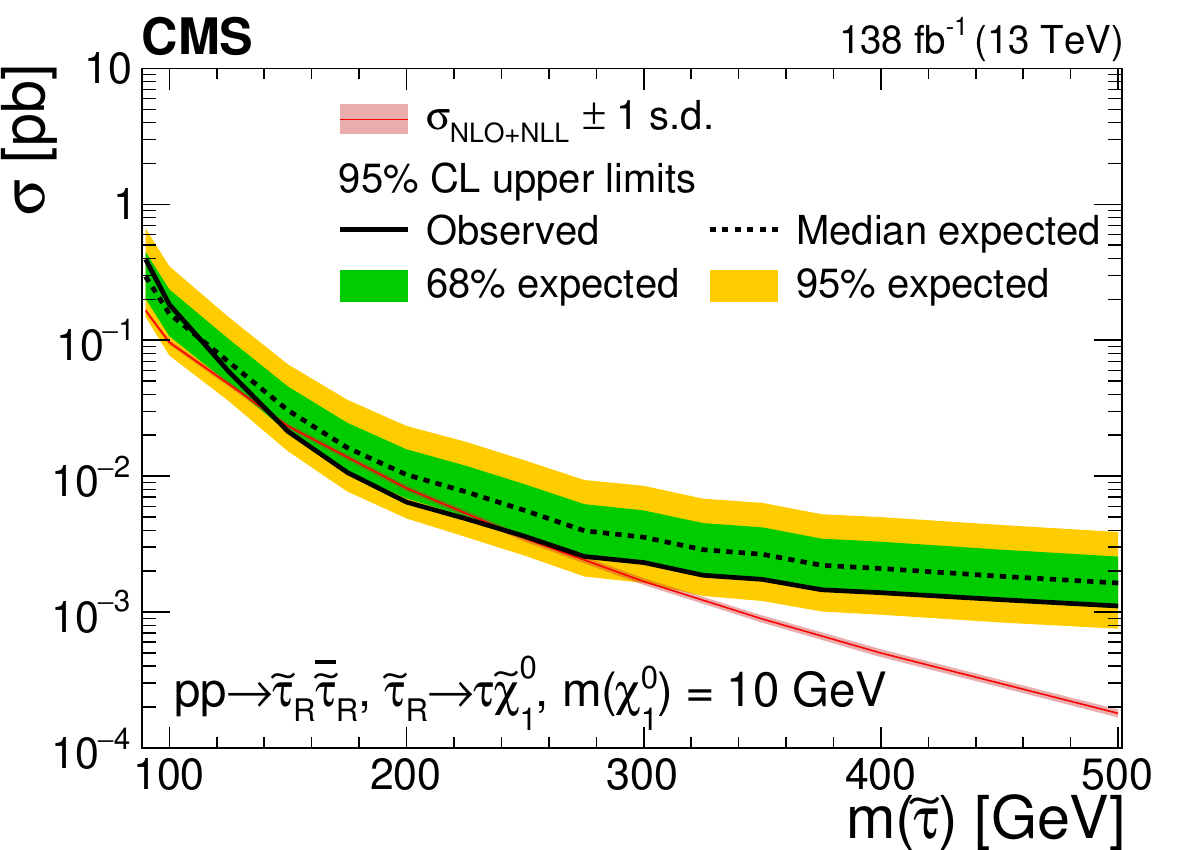} \\
\includegraphics[width=0.48\textwidth]{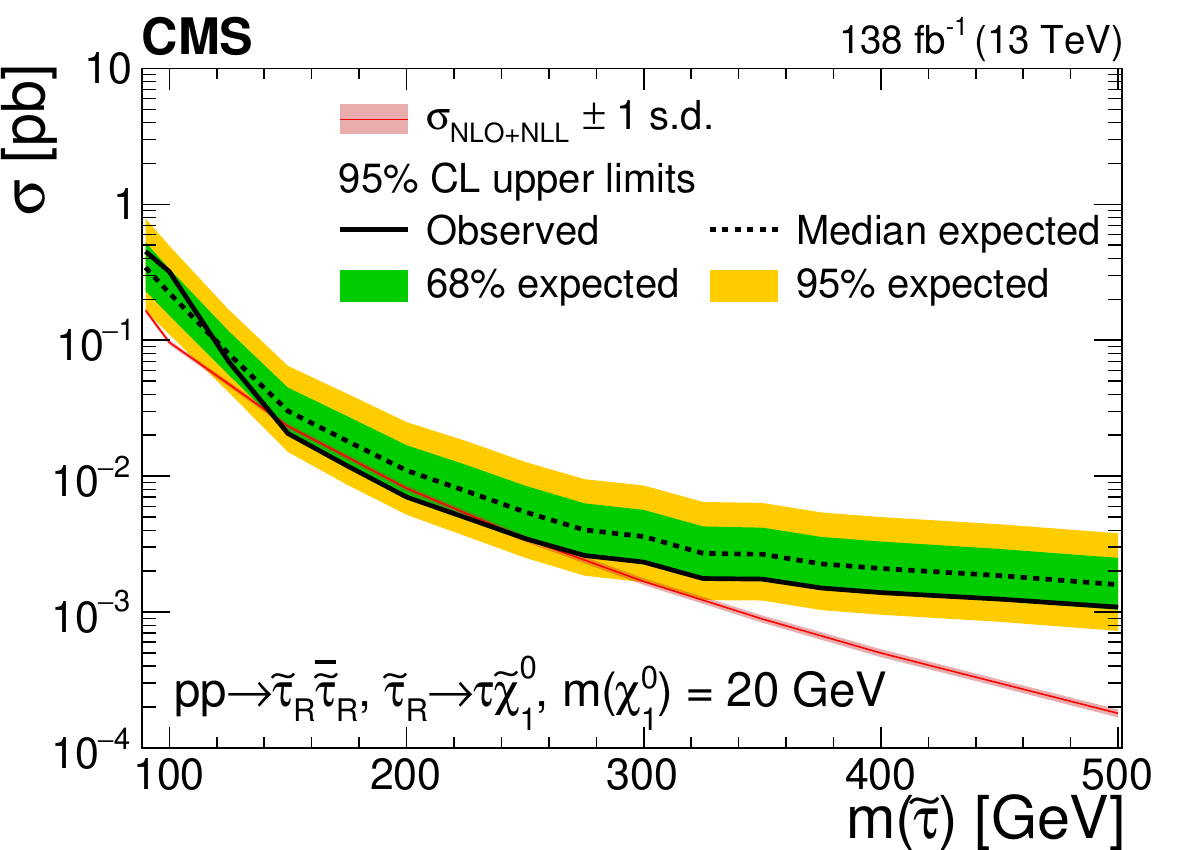}
\includegraphics[width=0.48\textwidth]{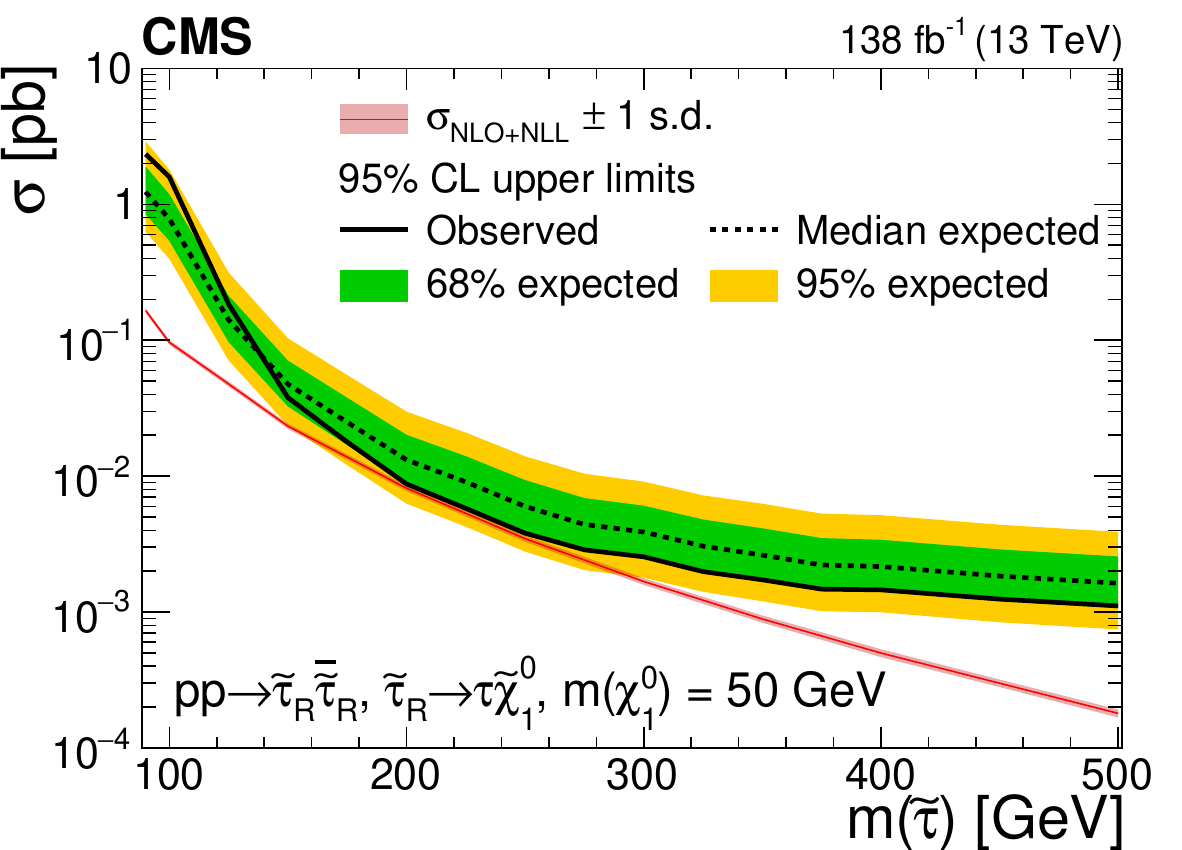}
\caption{Expected and observed 95\% \CL cross section upper limits as functions of the \sTau mass in the purely right-handed \sTau scenario for \lsp masses of 1, 10, 20, and 50\GeV (upper left to lower right). The inner (green) band and the outer (yellow) band indicate the regions containing 68 and 95\%, respectively, of the distribution of limits expected under the background-only hypothesis. The red line and thin shaded band indicate the NLO+NLL prediction for the signal production cross section calculated with \textsc{Resummino}~\cite{Fuks:2013lya}, and its uncertainty.}
\label{fig:limitsright}
\end{figure*}

\begin{figure*}[htbp!]
\centering
\includegraphics[width=0.48\textwidth]{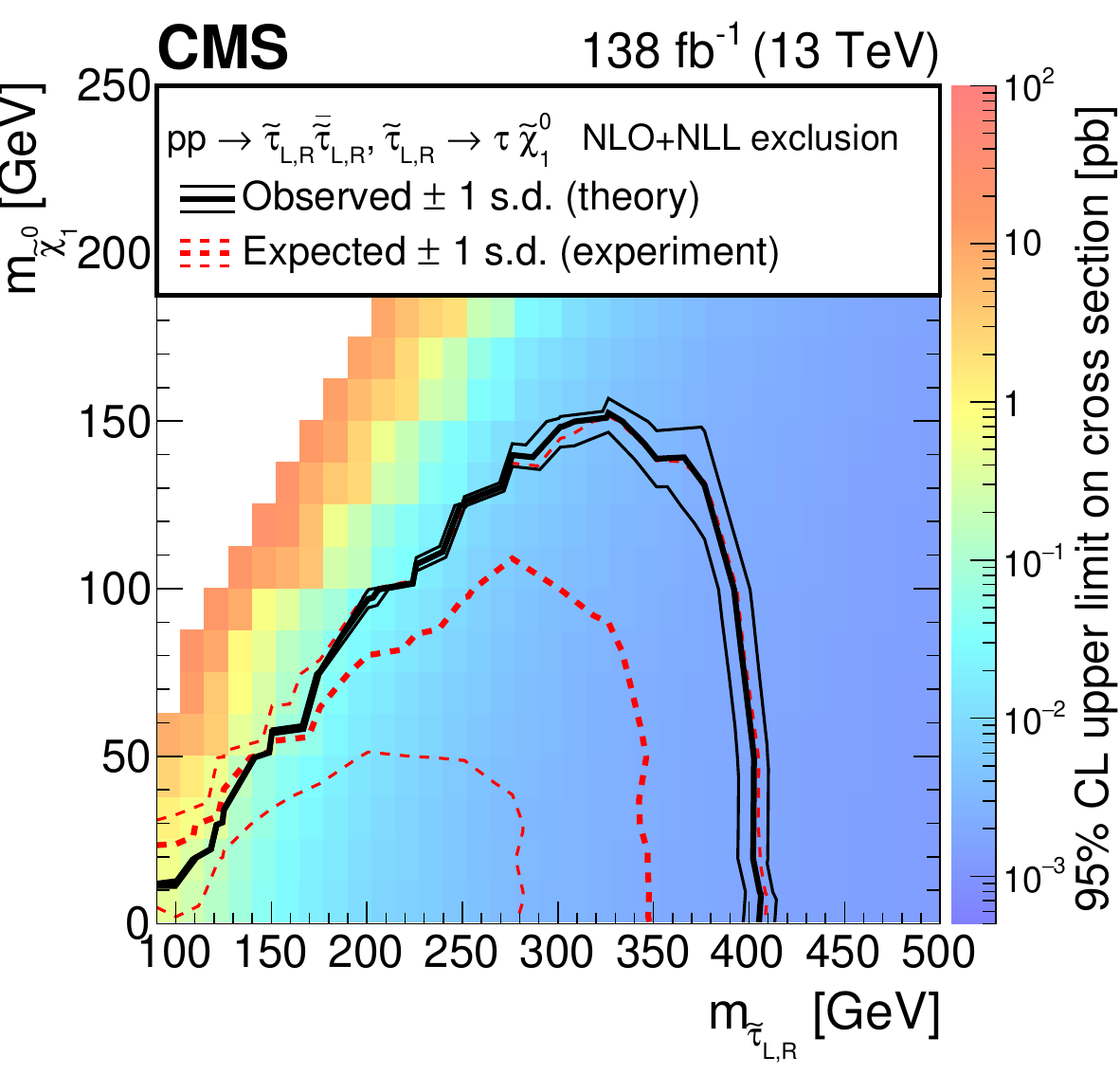}
\includegraphics[width=0.48\textwidth]{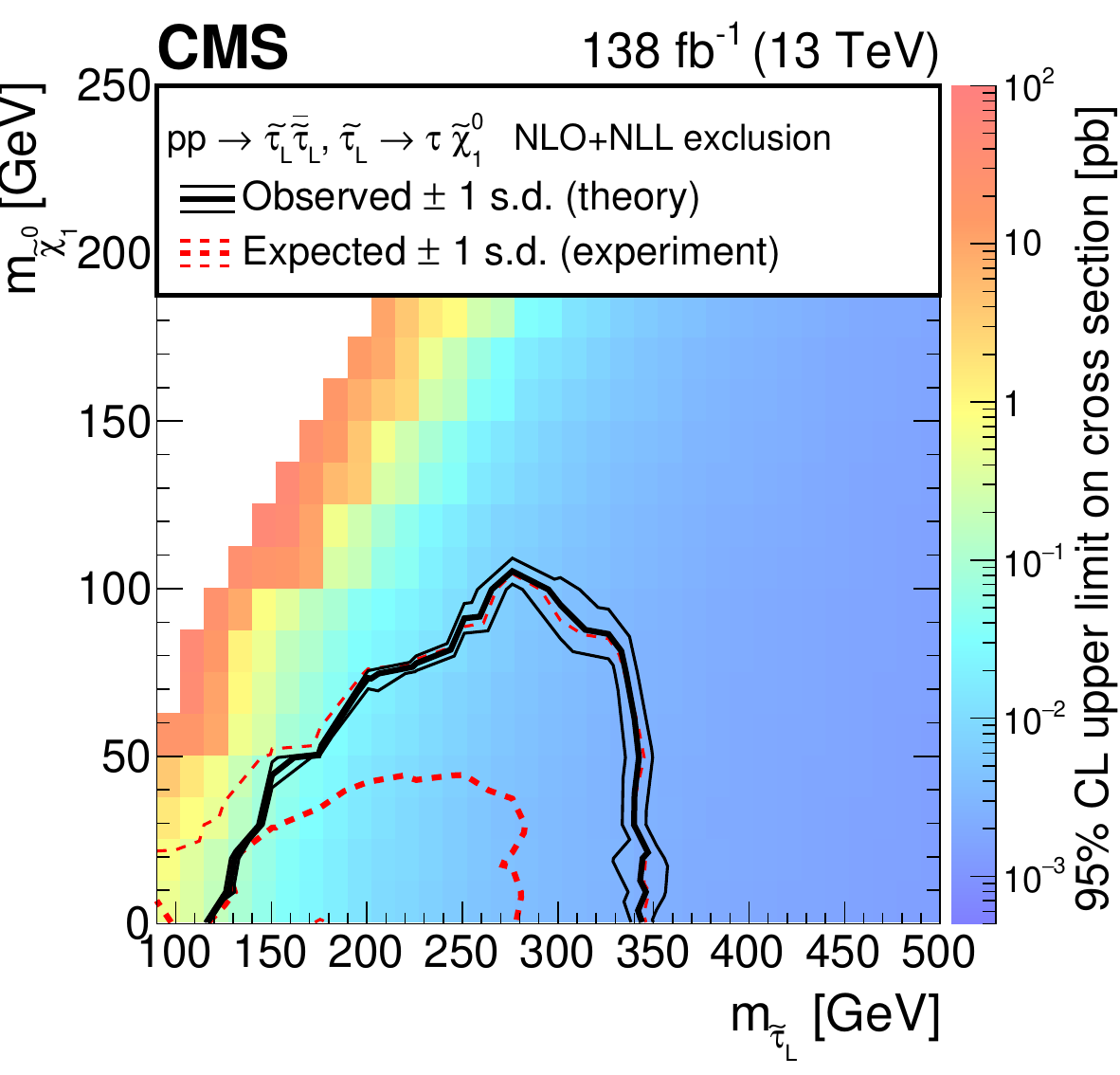}
\caption{Upper limits at 95\% \CL on the cross section for degenerate (left) and purely left-handed (right) \sTau pair production in the $m(\sTau)$-$m(\lsp)$ plane. The thick black (red) curves show the observed (expected) exclusion limits assuming NLO+NLL predictions for the signal cross sections. The thin black curves represent the variations in the observed limits obtained when these cross sections are varied by their $\pm1$ standard deviation uncertainties. The thin dashed red curves indicate the region containing 68\% of the distribution of limits expected under the background-only hypothesis.}
\label{fig:excllimits}
\end{figure*}

Expected and observed 95\% \CL cross section upper limits as a function of the \sTau mass for 4 choices of LSP mass are shown in Fig.~\ref{fig:limitsdegen} for \sTau pair production with promptly decaying \PGt sleptons in the degenerate scenario, in which we assume that both left- and right-handed \PGt sleptons are produced with the same mass, and in Figs.~\ref{fig:limitsleft} and~\ref{fig:limitsright}, in the purely left- and right-handed scenarios, respectively. Also shown is the theoretical prediction for the cross section of \sTau pair production in each scenario, as a function of the \sTau mass. In general, the cross section limits become less stringent for higher values of the \PSGczDo mass as a result of smaller experimental acceptance, caused in particular by the decreasing probability of the \tauh candidate to exceed the 40\GeV $\pt^{\tauh}$ threshold. Exclusion limits in the \sTau vs \PSGczDo mass plane are presented in Fig.~\ref{fig:excllimits} for promptly decaying \PGt sleptons in the degenerate and purely left-handed scenarios. In the degenerate scenarios \sTau masses up to 400\GeV are excluded at 95\% \CL under the hypothesis of a nearly massless LSP, while in the purely-left handed scenario, \sTau masses between 115 and 340\GeV are excluded under the same hypothesis. For values of the \sTau mass above $\approx$175--200\GeV, we generally observe exclusion limits that are approximately 1-standard deviation stronger than the expected exclusion limits, with the difference being largely driven by the deficits observed in SR bins 17 and 18. Exclusion limits for the purely right-handed scenario are not shown, because our sensitivity in this scenario is not yet sufficient to exclude a sizeable region in the \sTau vs \PSGczDo mass plane.

\begin{figure*}[ph!]
\centering
\includegraphics[width=0.45\textwidth]{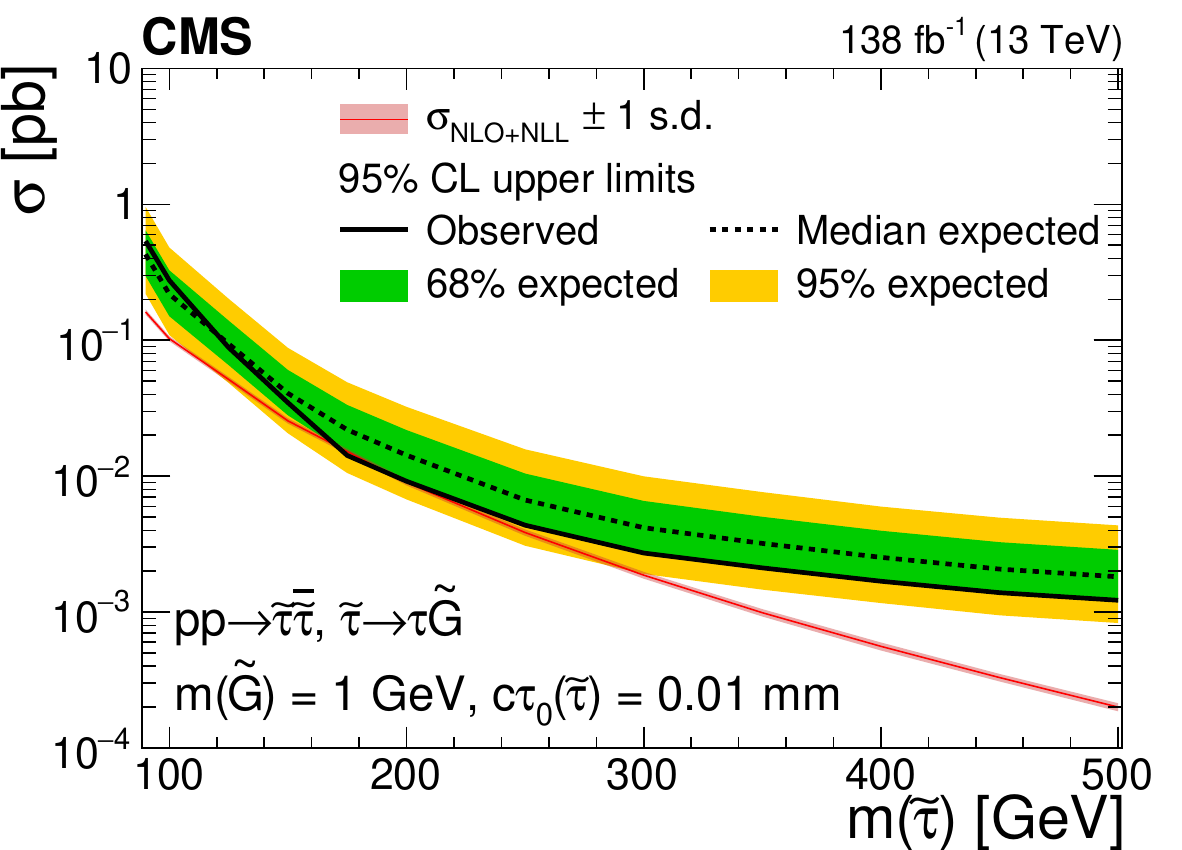}
\includegraphics[width=0.45\textwidth]{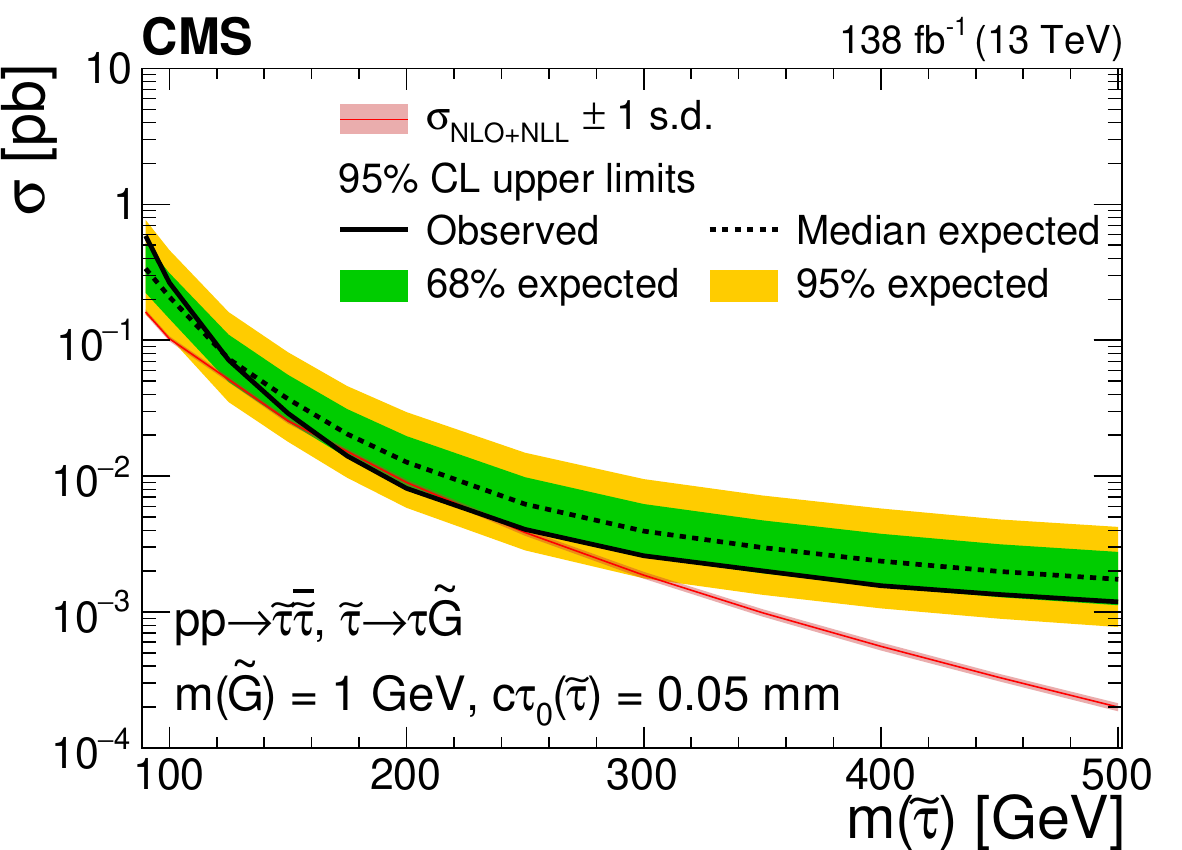} \\
\includegraphics[width=0.45\textwidth]{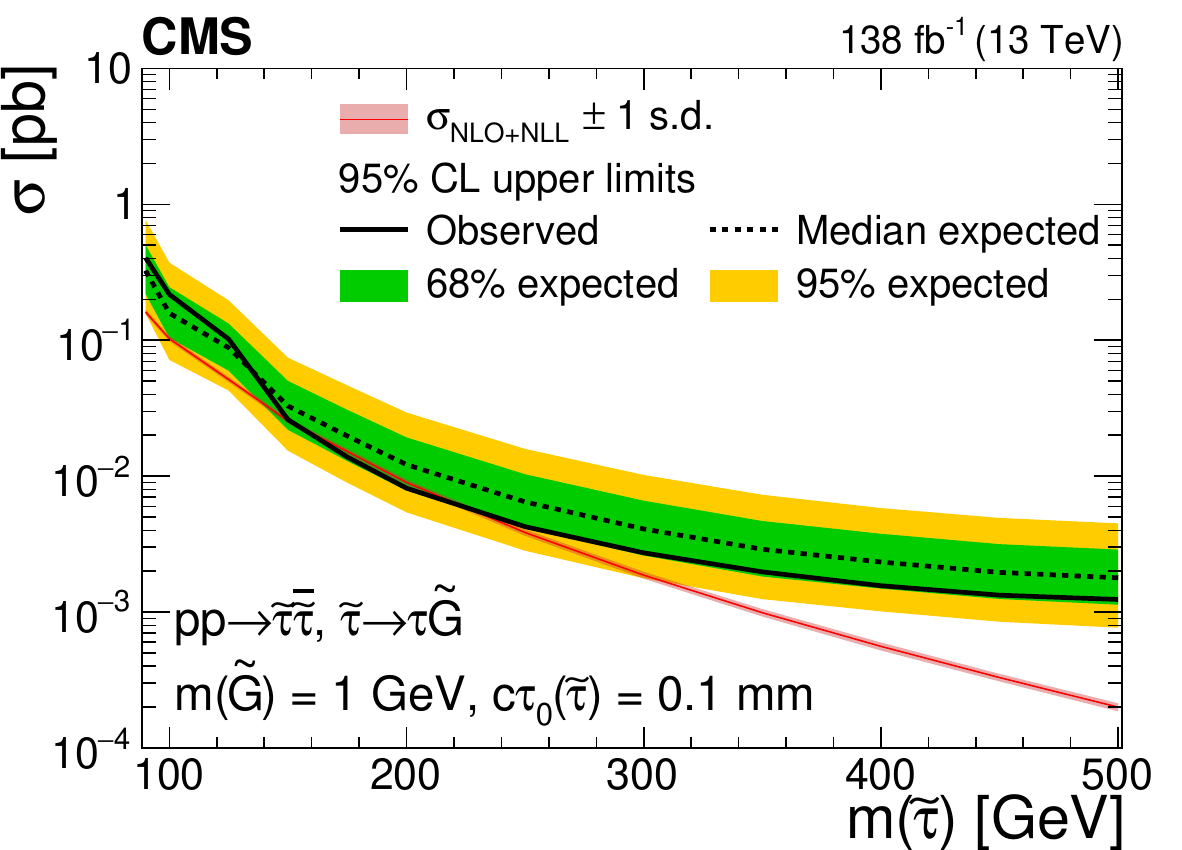}
\includegraphics[width=0.45\textwidth]{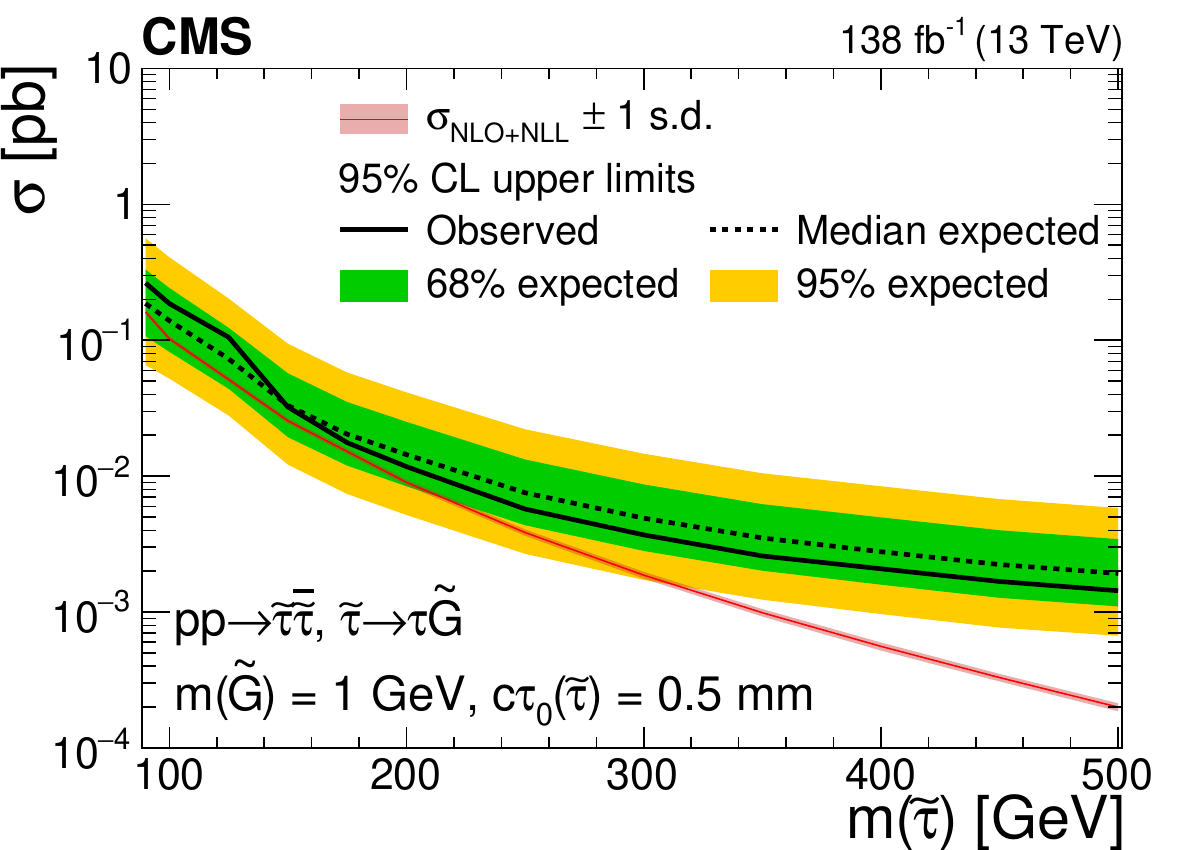} \\
\includegraphics[width=0.45\textwidth]{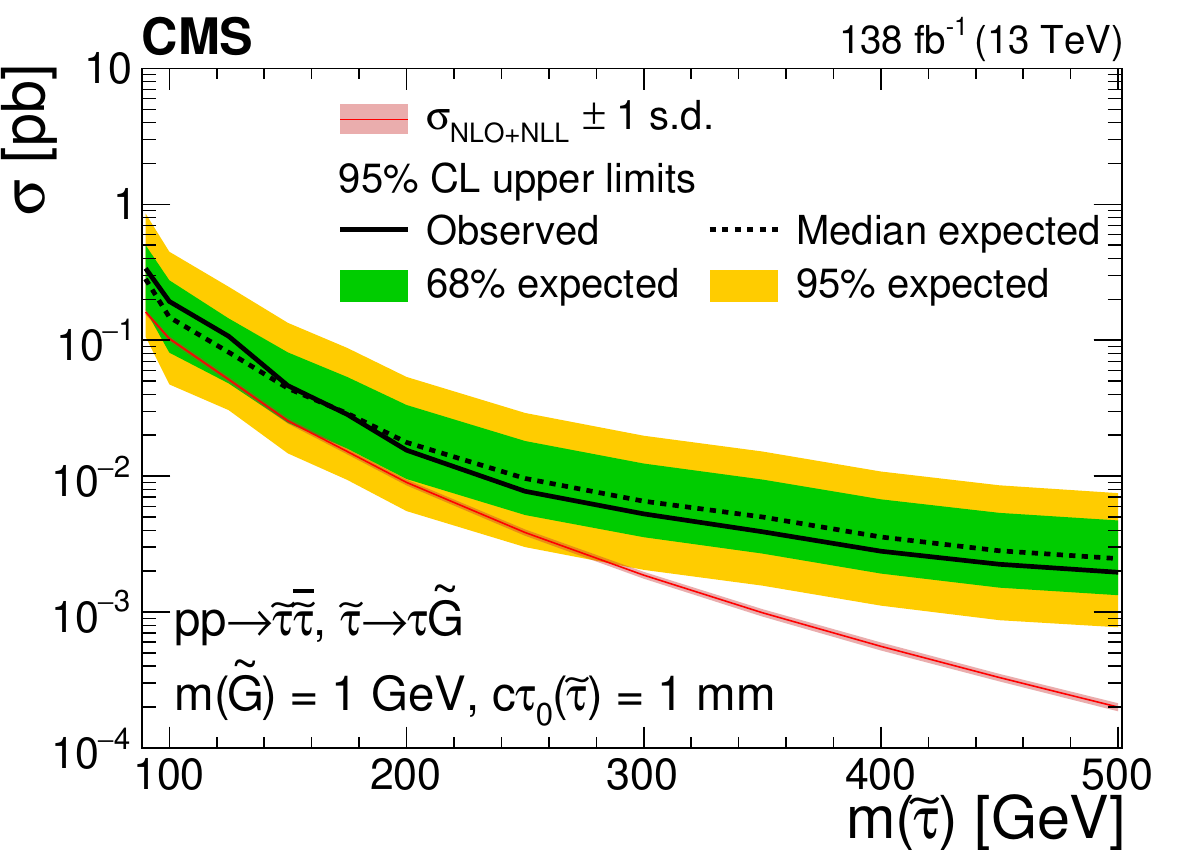}
\includegraphics[width=0.45\textwidth]{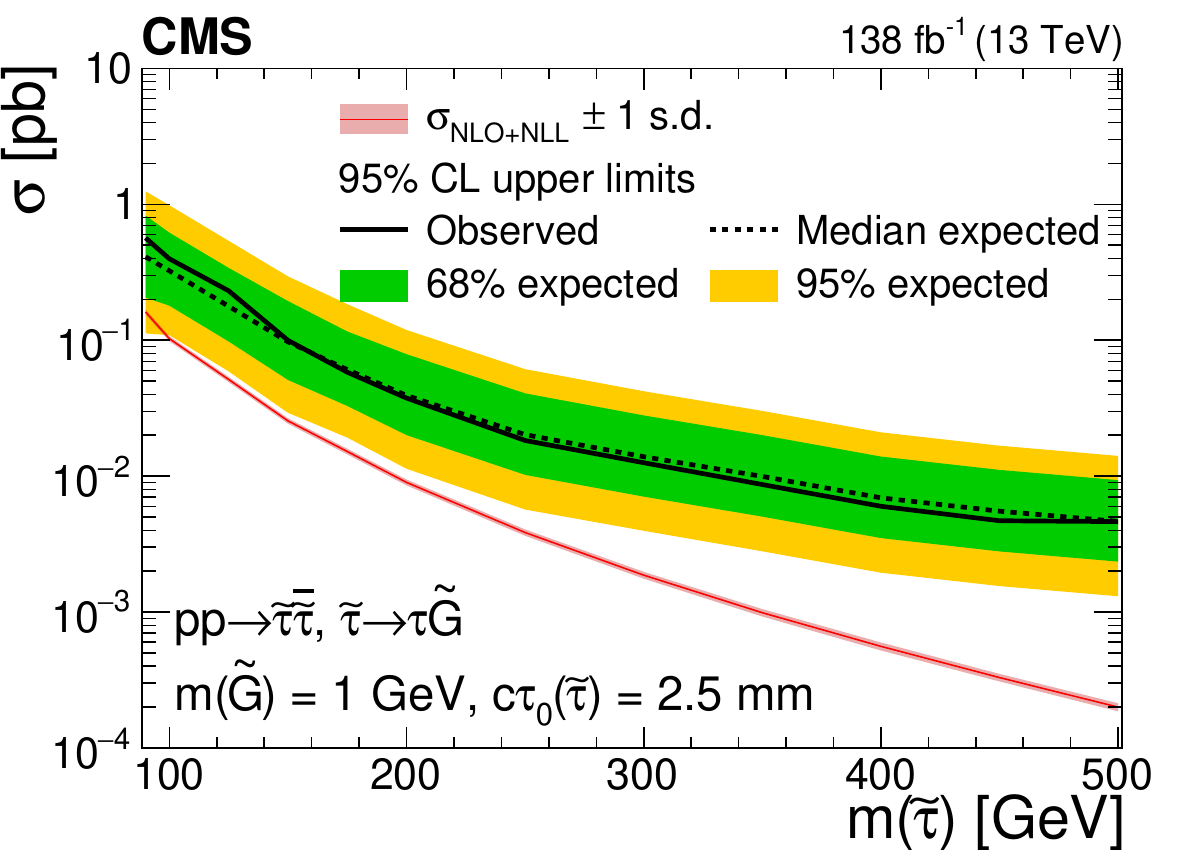}
\caption{Expected and observed 95\% \CL cross section upper limits as functions of the \sTau mass for long-lived \sTau in the maximally mixed scenario for an LSP mass of 1\GeV, and for \ctau values of 0.01, 0.05, 0.1, 0.5, 1, and 2.5\unit{mm} (upper left to lower right). The inner (green) band and the outer (yellow) band indicate the regions containing 68 and 95\%, respectively, of the distribution of limits expected under the background-only hypothesis. The red line and thin shaded band indicate the NLO+NLL prediction for the signal production cross section calculated with \textsc{Resummino}~\cite{Fuks:2013lya}, and its uncertainty.}
\label{fig:limitsll}
\end{figure*}

Expected and observed 95\% \CL cross section upper limits are shown in Fig.~\ref{fig:limitsll} for long-lived \PGt sleptons in the maximally mixed scenario, under the hypothesis of a nearly massless LSP. For a \sTau lifetime corresponding to $\ctau=0.1\unit{mm}$, \sTau masses between 150 and 220\GeV are excluded in this scenario, with the mass limits being assessed from the intersection of the observed exclusion curve with the theoretical prediction. The search sensitivity is reduced for longer \sTau lifetimes because the \textsc{DeepTau} algorithm is currently optimized for promptly produced \tauh candidates and is less efficient for more significantly displaced \tauh candidates.

\clearpage

\section {Summary}
\label{sec:summary}
A search for direct \PGt slepton (\PSGt) pair production has been performed in proton-proton collisions at a center-of-mass energy of 13\TeV in events with two hadronically decaying \PGt leptons and significant missing transverse momentum. The data used for this search correspond to an integrated luminosity of 138\fbinv collected in 2016--2018 with the CMS detector. Both prompt and displaced decays of the \PGt slepton are studied. Thirty-one different search regions are used in the analysis, based on kinematic observables that exploit expected differences between signal and background. No significant excess of events above the expected standard model background has been observed. Upper limits have been set on the cross section for direct \PSGt pair production for simplified models in which each \PSGt decays to a \PGt lepton and the lightest supersymmetric particle (LSP). For purely left-handed \sTau pair production with prompt decays, \sTau masses between 115 and 340\GeV are excluded at 95\% confidence level for the case of a nearly massless LSP, while for the degenerate production of left- and right-handed \sTau pairs, \sTau masses up to 400\GeV are excluded under the same hypothesis. The limits observed are the most stringent obtained thus far in the case of direct \sTau pair production with prompt \sTau decays, for both the purely left-handed and degenerate production scenarios. They represent a considerable improvement in sensitivity with respect to the previous CMS search reported in Ref.~\cite{CMS:2019eln}. In the context of long-lived \PGt sleptons, final states with displaced \tauh candidates are investigated for the first time. In a scenario with $\ctau(\sTau)=0.1\unit{mm}$, where $\tau_{0}$ denotes the mean proper lifetime of the \sTau, masses between 150 and 220\GeV are excluded for the case that the LSP is nearly massless.

\begin{acknowledgments}
We congratulate our colleagues in the CERN accelerator departments for the excellent performance of the LHC and thank the technical and administrative staffs at CERN and at other CMS institutes for their contributions to the success of the CMS effort. In addition, we gratefully acknowledge the computing centers and personnel of the Worldwide LHC Computing Grid and other centers for delivering so effectively the computing infrastructure essential to our analyses. Finally, we acknowledge the enduring support for the construction and operation of the LHC, the CMS detector, and the supporting computing infrastructure provided by the following funding agencies: BMBWF and FWF (Austria); FNRS and FWO (Belgium); CNPq, CAPES, FAPERJ, FAPERGS, and FAPESP (Brazil); MES and BNSF (Bulgaria); CERN; CAS, MoST, and NSFC (China); MINCIENCIAS (Colombia); MSES and CSF (Croatia); RIF (Cyprus); SENESCYT (Ecuador); MoER, ERC PUT and ERDF (Estonia); Academy of Finland, MEC, and HIP (Finland); CEA and CNRS/IN2P3 (France); BMBF, DFG, and HGF (Germany); GSRI (Greece); NKFIH (Hungary); DAE and DST (India); IPM (Iran); SFI (Ireland); INFN (Italy); MSIP and NRF (Republic of Korea); MES (Latvia); LAS (Lithuania); MOE and UM (Malaysia); BUAP, CINVESTAV, CONACYT, LNS, SEP, and UASLP-FAI (Mexico); MOS (Montenegro); MBIE (New Zealand); PAEC (Pakistan); MES and NSC (Poland); FCT (Portugal); MESTD (Serbia); MCIN/AEI and PCTI (Spain); MOSTR (Sri Lanka); Swiss Funding Agencies (Switzerland); MST (Taipei); MHESI and NSTDA (Thailand); TUBITAK and TENMAK (Turkey); NASU (Ukraine); STFC (United Kingdom); DOE and NSF (USA).

\hyphenation{Rachada-pisek} Individuals have received support from the Marie-Curie program and the European Research Council and Horizon 2020 Grant, contract Nos.\ 675440, 724704, 752730, 758316, 765710, 824093, 884104, and COST Action CA16108 (European Union); the Leventis Foundation; the Alfred P.\ Sloan Foundation; the Alexander von Humboldt Foundation; the Belgian Federal Science Policy Office; the Fonds pour la Formation \`a la Recherche dans l'Industrie et dans l'Agriculture (FRIA-Belgium); the Agentschap voor Innovatie door Wetenschap en Technologie (IWT-Belgium); the F.R.S.-FNRS and FWO (Belgium) under the ``Excellence of Science -- EOS" -- be.h project n.\ 30820817; the Beijing Municipal Science \& Technology Commission, No. Z191100007219010; the Ministry of Education, Youth and Sports (MEYS) of the Czech Republic; the Hellenic Foundation for Research and Innovation (HFRI), Project Number 2288 (Greece); the Deutsche Forschungsgemeinschaft (DFG), under Germany's Excellence Strategy -- EXC 2121 ``Quantum Universe" -- 390833306, and under project number 400140256 - GRK2497; the Hungarian Academy of Sciences, the New National Excellence Program - \'UNKP, the NKFIH research grants K 124845, K 124850, K 128713, K 128786, K 129058, K 131991, K 133046, K 138136, K 143460, K 143477, 2020-2.2.1-ED-2021-00181, and TKP2021-NKTA-64 (Hungary); the Council of Science and Industrial Research, India; the Latvian Council of Science; the Ministry of Education and Science, project no. 2022/WK/14, and the National Science Center, contracts Opus 2021/41/B/ST2/01369 and 2021/43/B/ST2/01552 (Poland); the Funda\c{c}\~ao para a Ci\^encia e a Tecnologia, grant CEECIND/01334/2018 (Portugal); the National Priorities Research Program by Qatar National Research Fund; MCIN/AEI/10.13039/501100011033, ERDF ``a way of making Europe", and the Programa Estatal de Fomento de la Investigaci{\'o}n Cient{\'i}fica y T{\'e}cnica de Excelencia Mar\'{\i}a de Maeztu, grant MDM-2017-0765 and Programa Severo Ochoa del Principado de Asturias (Spain); the Chulalongkorn Academic into Its 2nd Century Project Advancement Project, and the National Science, Research and Innovation Fund via the Program Management Unit for Human Resources \& Institutional Development, Research and Innovation, grant B05F650021 (Thailand); the Kavli Foundation; the Nvidia Corporation; the SuperMicro Corporation; the Welch Foundation, contract C-1845; and the Weston Havens Foundation (USA).
\end{acknowledgments}

\bibliography{auto_generated}

\providecommand{\href}[2]{#2}\begingroup\raggedright\begin{thebibliography}{10}%
\makeatletter
\providecommand{\hrefCMSnoop }[0]{\@secondoftwo}%
\makeatother
\providecommand{\doi}{\texttt{doi:}\begingroup \urlstyle{tt}\Url}

\bibitem{Ramond:1971gb}
\hrefCMSnoop {}{P.~Ramond, ``Dual theory for free fermions'',} \textit{ Phys.
  Rev. D} \textbf{ 3} (1971) 2415,
\href{http://dx.doi.org/10.1103/PhysRevD.3.2415}{\doi{10.1103/PhysRevD.3.2415}}.

\bibitem{Golfand:1971iw}
\href {http://www.jetpletters.ac.ru/ps/1584/article_24309.pdf}{Y.~A. Gol'fand
  and E.~P. Likhtman, ``Extension of the algebra of {P}oincar\'{e} group
  generators and violation of {P} invariance'',} \textit{ JETP Lett.} \textbf{
  13} (1971)
323.

\bibitem{Neveu:1971rx}
\hrefCMSnoop {}{A.~Neveu and J.~H. Schwarz, ``Factorizable dual model of
  pions'',} \textit{ Nucl. Phys. B} \textbf{ 31} (1971) 86,
\href{http://dx.doi.org/10.1016/0550-3213(71)90448-2}{\doi{10.1016/0550-3213(71)90448-2}}.

\bibitem{Volkov:1972jx}
\hrefCMSnoop {}{D.~V. Volkov and V.~P. Akulov, ``Possible universal neutrino
  interaction'',} \textit{ JETP Lett.} \textbf{ 16} (1972) 438,
\href{http://dx.doi.org/10.1007/BFb0105270}{\doi{10.1007/BFb0105270}}.

\bibitem{Wess:1973kz}
\hrefCMSnoop {}{J.~Wess and B.~Zumino, ``A {Lagrangian} model invariant under
  supergauge transformations'',} \textit{ Phys. Lett. B} \textbf{ 49} (1974)
  52,
\href{http://dx.doi.org/10.1016/0370-2693(74)90578-4}{\doi{10.1016/0370-2693(74)90578-4}}.

\bibitem{Wess:1974tw}
\hrefCMSnoop {}{J.~Wess and B.~Zumino, ``Supergauge transformations in four
  dimensions'',} \textit{ Nucl. Phys. B} \textbf{ 70} (1974) 39,
\href{http://dx.doi.org/10.1016/0550-3213(74)90355-1}{\doi{10.1016/0550-3213(74)90355-1}}.

\bibitem{Fayet:1974pd}
\hrefCMSnoop {}{P.~Fayet, ``Supergauge invariant extension of the {H}iggs
  mechanism and a model for the electron and its neutrino'',} \textit{ Nucl.
  Phys. B} \textbf{ 90} (1975) 104,
\href{http://dx.doi.org/10.1016/0550-3213(75)90636-7}{\doi{10.1016/0550-3213(75)90636-7}}.

\bibitem{Nilles:1983ge}
\hrefCMSnoop {}{H.~P. Nilles, ``Supersymmetry, supergravity and particle
  physics'',} \textit{ Phys. Rep.} \textbf{ 110} (1984) 1,
\href{http://dx.doi.org/10.1016/0370-1573(84)90008-5}{\doi{10.1016/0370-1573(84)90008-5}}.

\bibitem{Gildener:1976ai}
\hrefCMSnoop {}{E.~Gildener, ``Gauge symmetry hierarchies'',} \textit{ Phys.
  Rev. D} \textbf{ 14} (1976) 1667,
\href{http://dx.doi.org/10.1103/PhysRevD.14.1667}{\doi{10.1103/PhysRevD.14.1667}}.

\bibitem{Veltman:1976rt}
\hrefCMSnoop {}{M.~J.~G. Veltman, ``Second threshold in weak interactions'',}
  \textit{ Acta Phys. Polon. B} \textbf{ 8} (1977) 475,
\href{http://dx.doi.org/10.1016/B978-0-444-88807-5.50087-X}{\doi{10.1016/B978-0-444-88807-5.50087-X}}.

\bibitem{'tHooft:1979bh}
\hrefCMSnoop {}{G.~'t~Hooft, ``Naturalness, chiral symmetry, and spontaneous
  chiral symmetry breaking'',} \textit{ NATO Sci. Ser. B} \textbf{ 59} (1980)
135.

\bibitem{Witten:1981nf}
\hrefCMSnoop {}{E.~Witten, ``Dynamical breaking of supersymmetry'',} \textit{
  Nucl. Phys. B} \textbf{ 188} (1981) 513,
\href{http://dx.doi.org/10.1016/0550-3213(81)90006-7}{\doi{10.1016/0550-3213(81)90006-7}}.

\bibitem{Farrar:1978xj}
\hrefCMSnoop {}{G.~R. Farrar and P.~Fayet, ``Phenomenology of the production,
  decay, and detection of new hadronic states associated with supersymmetry'',}
  \textit{ Phys. Lett. B} \textbf{ 76} (1978) 575,
\href{http://dx.doi.org/10.1016/0370-2693(78)90858-4}{\doi{10.1016/0370-2693(78)90858-4}}.

\bibitem{Goldberg:1983nd}
\hrefCMSnoop {}{H.~Goldberg, ``Constraint on the photino mass from
  cosmology'',} \textit{ Phys. Rev. Lett.} \textbf{ 50} (1983) 1419,
  \href{http://dx.doi.org/10.1103/PhysRevLett.50.1419}{\doi{10.1103/PhysRevLett.50.1419}}.
[Erratum: \DOI{10.1103/PhysRevLett.103.099905}].

\bibitem{Ellis:1983ew}
J.~R. Ellis\hrefCMSnoop {}{ {et~al.}, ``Supersymmetric relics from the {Big
  Bang}'',} \textit{ Nucl. Phys. B} \textbf{ 238} (1984) 453,
\href{http://dx.doi.org/10.1016/0550-3213(84)90461-9}{\doi{10.1016/0550-3213(84)90461-9}}.

\bibitem{darkmatter}
\hrefCMSnoop {}{G.~Jungman, M.~Kamionkowski, and K.~Griest, ``Supersymmetric
  dark matter'',} \textit{ Phys. Rept.} \textbf{ 267} (1996) 195,
  \href{http://dx.doi.org/10.1016/0370-1573(95)00058-5}{\doi{10.1016/0370-1573(95)00058-5}},
\href{http://www.arXiv.org/abs/hep-ph/9506380}{\texttt{arXiv:hep-ph/9506380}}.

\bibitem{WMAP}
\hrefCMSnoop {}{G.~Hinshaw {et~al.}, ``Nine-year {Wilkinson Microwave
  Anisotropy Probe (WMAP)} observations: cosmological parameter results'',}
  \textit{ Astrophys. J. Suppl.} \textbf{ 208} (2013) 19,
  \href{http://dx.doi.org/10.1088/0067-0049/208/2/19}{\doi{10.1088/0067-0049/208/2/19}},
\href{http://www.arXiv.org/abs/1212.5226}{\texttt{arXiv:1212.5226}}.

\bibitem{Griest:1990kh}
\hrefCMSnoop {}{K.~Griest and D.~Seckel, ``Three exceptions in the calculation
  of relic abundances'',} \textit{ Phys. Rev. D} \textbf{ 43} (1991) 3191,
\href{http://dx.doi.org/10.1103/PhysRevD.43.3191}{\doi{10.1103/PhysRevD.43.3191}}.

\bibitem{Vasquez}
\hrefCMSnoop {}{D.~Albornoz~V{\'a}squez, G.~B{\'e}langer, and C.~B{\oe}hm,
  ``Revisiting light neutralino scenarios in the {MSSM}'',} \textit{ Phys. Rev.
  D} \textbf{ 84} (2011) 095015,
  \href{http://dx.doi.org/10.1103/PhysRevD.84.095015}{\doi{10.1103/PhysRevD.84.095015}},
\href{http://www.arXiv.org/abs/1108.1338}{\texttt{arXiv:1108.1338}}.

\bibitem{King}
\hrefCMSnoop {}{S.~F. King, J.~P. Roberts, and D.~P. Roy, ``Natural dark matter
  in {SUSY GUTs} with non-universal gaugino masses'',} \textit{ JHEP} \textbf{
  10} (2007) 106,
  \href{http://dx.doi.org/10.1088/1126-6708/2007/10/106}{\doi{10.1088/1126-6708/2007/10/106}},
\href{http://www.arXiv.org/abs/0705.4219}{\texttt{arXiv:0705.4219}}.

\bibitem{Battaglia:2001zp}
M.~Battaglia\hrefCMSnoop {}{ {et~al.}, ``Proposed post-{LEP} benchmarks for
  supersymmetry'',} \textit{ Eur. Phys. J. C} \textbf{ 22} (2001) 535,
  \href{http://dx.doi.org/10.1007/s100520100792}{\doi{10.1007/s100520100792}},
\href{http://www.arXiv.org/abs/hep-ph/0106204}{\texttt{arXiv:hep-ph/0106204}}.

\bibitem{Arnowitt:2008bz}
R.~L. Arnowitt\hrefCMSnoop {}{ {et~al.}, ``Determining the dark matter relic
  density in the minimal supergravity stau-neutralino coannihilation region at
  the {Large Hadron Collider}'',} \textit{ Phys. Rev. Lett.} \textbf{ 100}
  (2008) 231802,
  \href{http://dx.doi.org/10.1103/PhysRevLett.100.231802}{\doi{10.1103/PhysRevLett.100.231802}},
\href{http://www.arXiv.org/abs/0802.2968}{\texttt{arXiv:0802.2968}}.

\bibitem{Belanger:2012jn}
\hrefCMSnoop {}{G.~B{\'e}langer, S.~Biswas, C.~Boehm, and B.~Mukhopadhyaya,
  ``Light neutralino dark matter in the {MSSM} and its implication for {LHC}
  searches for staus'',} \textit{ JHEP} \textbf{ 12} (2012) 076,
  \href{http://dx.doi.org/10.1007/JHEP12(2012)076}{\doi{10.1007/JHEP12(2012)076}},
\href{http://www.arXiv.org/abs/1206.5404}{\texttt{arXiv:1206.5404}}.

\bibitem{Arganda:2018hdn}
\hrefCMSnoop {}{E.~Arganda, V.~Martin-Lozano, A.~D. Medina, and N.~Mileo,
  ``Potential discovery of staus through heavy {H}iggs boson decays at the
  {LHC}'',} \textit{ JHEP} \textbf{ 09} (2018) 056,
  \href{http://dx.doi.org/10.1007/JHEP09(2018)056}{\doi{10.1007/JHEP09(2018)056}},
\href{http://www.arXiv.org/abs/1804.10698}{\texttt{arXiv:1804.10698}}.

\bibitem{Arganda:2021qgi}
\hrefCMSnoop {}{E.~Arganda, V.~Mart{\'\i}n-Lozano, A.~D. Medina, and N.~I.
  Mileo, ``Discovery and exclusion prospects for staus produced by heavy
  {H}iggs boson decays at the {LHC}'',} \textit{ Adv. High Energy Phys.}
  \textbf{ 2022} (2022) 2569290,
  \href{http://dx.doi.org/10.1155/2022/2569290}{\doi{10.1155/2022/2569290}},
  \href{http://www.arXiv.org/abs/2102.02290}{\texttt{arXiv:2102.02290}}.

\bibitem{Evans:2016zau}
\hrefCMSnoop {}{J.~A. Evans and J.~Shelton, ``Long-lived staus and displaced
  leptons at the {LHC}'',} \textit{ JHEP} \textbf{ 04} (2016) 056,
  \href{http://dx.doi.org/10.1007/JHEP04(2016)056}{\doi{10.1007/JHEP04(2016)056}},
  \href{http://www.arXiv.org/abs/1601.01326}{\texttt{arXiv:1601.01326}}.

\bibitem{Alwall:2008ag}
\hrefCMSnoop {}{J.~Alwall, P.~C. Schuster, and N.~Toro, ``Simplified models for
  a first characterization of new physics at the {LHC}'',} \textit{ Phys. Rev.
  D} \textbf{ 79} (2009) 075020,
  \href{http://dx.doi.org/10.1103/PhysRevD.79.075020}{\doi{10.1103/PhysRevD.79.075020}},
\href{http://www.arXiv.org/abs/0810.3921}{\texttt{arXiv:0810.3921}}.

\bibitem{Alwall:2008va}
\hrefCMSnoop {}{J.~Alwall, M.-P. Le, M.~Lisanti, and J.~G. Wacker,
  ``Model-independent jets plus missing energy searches'',} \textit{ Phys. Rev.
  D} \textbf{ 79} (2009) 015005,
  \href{http://dx.doi.org/10.1103/PhysRevD.79.015005}{\doi{10.1103/PhysRevD.79.015005}},
\href{http://www.arXiv.org/abs/0809.3264}{\texttt{arXiv:0809.3264}}.

\bibitem{Alves:2011wf}
\hrefCMSnoop {}{{LHC New Physics Working Group}, ``Simplified models for {LHC}
  new physics searches'',} \textit{ J. Phys. G} \textbf{ 39} (2012) 105005,
  \href{http://dx.doi.org/10.1088/0954-3899/39/10/105005}{\doi{10.1088/0954-3899/39/10/105005}},
\href{http://www.arXiv.org/abs/1105.2838}{\texttt{arXiv:1105.2838}}.

\bibitem{Fuks:2013lya}
\hrefCMSnoop {}{B.~Fuks, M.~Klasen, D.~R. Lamprea, and M.~Rothering,
  ``Revisiting slepton pair production at the {Large Hadron Collider}'',}
  \textit{ JHEP} \textbf{ 01} (2014) 168,
  \href{http://dx.doi.org/10.1007/JHEP01(2014)168}{\doi{10.1007/JHEP01(2014)168}},
\href{http://www.arXiv.org/abs/1310.2621}{\texttt{arXiv:1310.2621}}.

\bibitem{Heister:2001nk}
\hrefCMSnoop {}{{ALEPH} Collaboration, ``Search for scalar leptons in
  e$^{+}$e$^{-}$ collisions at center-of-mass energies up to 209 {GeV}'',}
  \textit{ Phys. Lett. B} \textbf{ 526} (2002) 206,
  \href{http://dx.doi.org/10.1016/S0370-2693(01)01494-0}{\doi{10.1016/S0370-2693(01)01494-0}},
\href{http://www.arXiv.org/abs/hep-ex/0112011}{\texttt{arXiv:hep-ex/0112011}}.

\bibitem{Abdallah:2003xe}
\hrefCMSnoop {}{{DELPHI} Collaboration, ``Searches for supersymmetric particles
  in e$^{+}$e$^{-}$ collisions up to 208 {GeV} and interpretation of the
  results within the {MSSM}'',} \textit{ Eur. Phys. J. C} \textbf{ 31} (2003)
  421,
  \href{http://dx.doi.org/10.1140/epjc/s2003-01355-5}{\doi{10.1140/epjc/s2003-01355-5}},
\href{http://www.arXiv.org/abs/hep-ex/0311019}{\texttt{arXiv:hep-ex/0311019}}.

\bibitem{Achard:2003ge}
\hrefCMSnoop {}{{L3} Collaboration, ``Search for scalar leptons and scalar
  quarks at {LEP}'',} \textit{ Phys. Lett. B} \textbf{ 580} (2004) 37,
  \href{http://dx.doi.org/10.1016/j.physletb.2003.10.010}{\doi{10.1016/j.physletb.2003.10.010}},
\href{http://www.arXiv.org/abs/hep-ex/0310007}{\texttt{arXiv:hep-ex/0310007}}.

\bibitem{Abbiendi:2003ji}
\hrefCMSnoop {}{{OPAL} Collaboration, ``Search for anomalous production of
  dilepton events with missing transverse momentum in e$^{+}$e$^{-}$ collisions
  at $\sqrt{s}$ = 183 {GeV} to 209 {GeV}'',} \textit{ Eur. Phys. J. C} \textbf{
  32} (2004) 453,
  \href{http://dx.doi.org/10.1140/epjc/s2003-01466-y}{\doi{10.1140/epjc/s2003-01466-y}},
\href{http://www.arXiv.org/abs/hep-ex/0309014}{\texttt{arXiv:hep-ex/0309014}}.

\bibitem{Aad:2014yka}
\hrefCMSnoop {}{{ATLAS Collaboration}, ``Search for the direct production of
  charginos, neutralinos and staus in final states with at least two
  hadronically decaying taus and missing transverse momentum in pp collisions
  at $\sqrt{s}$ = 8{\TeV} with the {ATLAS} detector'',} \textit{ JHEP} \textbf{
  10} (2014) 96,
  \href{http://dx.doi.org/10.1007/JHEP10(2014)096}{\doi{10.1007/JHEP10(2014)096}},
\href{http://www.arXiv.org/abs/1407.0350}{\texttt{arXiv:1407.0350}}.

\bibitem{Aad:2015eda}
\hrefCMSnoop {}{{ATLAS Collaboration}, ``Search for the electroweak production
  of supersymmetric particles in $\sqrt{s} = 8$ {TeV} pp collisions with the
  {ATLAS} detector'',} \textit{ Phys. Rev. D} \textbf{ 93} (2016) 052002,
  \href{http://dx.doi.org/10.1103/PhysRevD.93.052002}{\doi{10.1103/PhysRevD.93.052002}},
\href{http://www.arXiv.org/abs/1509.07152}{\texttt{arXiv:1509.07152}}.

\bibitem{SUS14022}
\hrefCMSnoop {}{{CMS Collaboration}, ``Search for electroweak production of
  charginos in final states with two tau leptons in pp collisions at $
  \sqrt{s}=8 $ {TeV}'',} \textit{ JHEP} \textbf{ 04} (2017) 018,
  \href{http://dx.doi.org/10.1007/JHEP04(2017)018}{\doi{10.1007/JHEP04(2017)018}},
\href{http://www.arXiv.org/abs/1610.04870}{\texttt{arXiv:1610.04870}}.

\bibitem{Aad:2019byo}
\hrefCMSnoop {}{{ATLAS Collaboration}, ``Search for direct stau production in
  events with two hadronic $\tau$-leptons in $\sqrt{s} = 13$ {TeV} $pp$
  collisions with the {ATLAS} detector'',} \textit{ Phys. Rev. D} \textbf{ 101}
  (2020) 032009,
  \href{http://dx.doi.org/10.1103/PhysRevD.101.032009}{\doi{10.1103/PhysRevD.101.032009}},
  \href{http://www.arXiv.org/abs/1911.06660}{\texttt{arXiv:1911.06660}}.

\bibitem{CMS:2019eln}
\hrefCMSnoop {}{{CMS Collaboration}, ``Search for direct pair production of
  supersymmetric partners to the $\tau$ lepton in proton-proton collisions at
  $\sqrt{s}=$ 13 {TeV}'',} \textit{ Eur. Phys. J. C} \textbf{ 80} (2020) 189,
  \href{http://dx.doi.org/10.1140/epjc/s10052-020-7739-7}{\doi{10.1140/epjc/s10052-020-7739-7}},
  \href{http://www.arXiv.org/abs/1907.13179}{\texttt{arXiv:1907.13179}}.

\bibitem{Abbiendi:2005gc}
\hrefCMSnoop {}{{OPAL} Collaboration, ``Searches for gauge-mediated
  supersymmetry breaking topologies in e$^{+}$e$^{-}$ collisions at
  centre-of-mass energies up to $\sqrt{s}$ = 209 {GeV}'',} \textit{ Eur. Phys.
  J. C} \textbf{ 46} (2006) 307,
  \href{http://dx.doi.org/10.1140/epjc/s2006-02524-8}{\doi{10.1140/epjc/s2006-02524-8}},
  \href{http://www.arXiv.org/abs/hep-ex/0507048}{\texttt{arXiv:hep-ex/0507048}}.

\bibitem{Aad:2020bay}
\hrefCMSnoop {}{{ATLAS Collaboration}, ``Search for displaced leptons in
  $\sqrt{s} = 13$ {TeV} $pp$ collisions with the {ATLAS} detector'',} \textit{
  Phys. Rev. Lett.} \textbf{ 127} (2021) 051802,
  \href{http://dx.doi.org/10.1103/PhysRevLett.127.051802}{\doi{10.1103/PhysRevLett.127.051802}},
  \href{http://www.arXiv.org/abs/2011.07812}{\texttt{arXiv:2011.07812}}.

\bibitem{Sirunyan:2019drn}
\hrefCMSnoop {}{{CMS Collaboration}, ``An embedding technique to determine
  $\tau\tau$ backgrounds in proton-proton collision data'',} \textit{ JINST}
  \textbf{ 14} (2019) P06032,
  \href{http://dx.doi.org/10.1088/1748-0221/14/06/P06032}{\doi{10.1088/1748-0221/14/06/P06032}},
  \href{http://www.arXiv.org/abs/1903.01216}{\texttt{arXiv:1903.01216}}.

\bibitem{CMS:2022prd}
\hrefCMSnoop {}{{CMS Collaboration}, ``Identification of hadronic tau lepton
  decays using a deep neural network'',} 2022.
  \href{http://www.arXiv.org/abs/2201.08458}{\texttt{arXiv:2201.08458}}.
  Submitted to {\it JINST}.

\bibitem{hepdata}
\hrefCMSnoop {}{}{HEPD}ata record for this analysis, 2022.
\newblock
  \href{http://dx.doi.org/10.17182/hepdata.131308}{\doi{10.17182/hepdata.131308}}.

\bibitem{CMS:2020cmk}
\hrefCMSnoop {}{{CMS Collaboration}, ``Performance of the {CMS} {L}evel-1
  trigger in proton-proton collisions at $\sqrt{s} = 13$\,{TeV}'',} \textit{
  JINST} \textbf{ 15} (2020) P10017,
  \href{http://dx.doi.org/10.1088/1748-0221/15/10/P10017}{\doi{10.1088/1748-0221/15/10/P10017}},
  \href{http://www.arXiv.org/abs/2006.10165}{\texttt{arXiv:2006.10165}}.

\bibitem{CMS:2016ngn}
\hrefCMSnoop {}{{CMS Collaboration}, ``The {CMS} trigger system'',} \textit{
  JINST} \textbf{ 12} (2017) P01020,
  \href{http://dx.doi.org/10.1088/1748-0221/12/01/P01020}{\doi{10.1088/1748-0221/12/01/P01020}},
\href{http://www.arXiv.org/abs/1609.02366}{\texttt{arXiv:1609.02366}}.

\bibitem{Sirunyan:2017ulk}
\hrefCMSnoop {}{{CMS Collaboration}, ``{Particle-flow reconstruction and global
  event description with the CMS detector}'',} \textit{ JINST} \textbf{ 12}
  (2017) P10003,
  \href{http://dx.doi.org/10.1088/1748-0221/12/10/P10003}{\doi{10.1088/1748-0221/12/10/P10003}},
\href{http://www.arXiv.org/abs/1706.04965}{\texttt{arXiv:1706.04965}}.

\bibitem{Sirunyan:2019kia}
\hrefCMSnoop {}{{CMS Collaboration}, ``Performance of missing transverse
  momentum reconstruction in proton-proton collisions at $\sqrt{s} =$ 13 {TeV}
  using the {CMS} detector'',} \textit{ JINST} \textbf{ 14} (2019) P07004,
  \href{http://dx.doi.org/10.1088/1748-0221/14/07/P07004}{\doi{10.1088/1748-0221/14/07/P07004}},
\href{http://www.arXiv.org/abs/1903.06078}{\texttt{arXiv:1903.06078}}.

\bibitem{CMS-TDR-15-02}
\href {http://cds.cern.ch/record/2020886}{{CMS Collaboration}, ``Technical
  proposal for the {Phase-II} upgrade of the {Compact Muon Solenoid}'',} CMS
  Technical Proposal CERN-LHCC-2015-010, CMS-TDR-15-02, 2015.

\bibitem{Cacciari:2008gp}
\hrefCMSnoop {}{M.~Cacciari, G.~P. Salam, and G.~Soyez, ``The anti-\kt jet
  clustering algorithm'',} \textit{ JHEP} \textbf{ 04} (2008) 063,
  \href{http://dx.doi.org/10.1088/1126-6708/2008/04/063}{\doi{10.1088/1126-6708/2008/04/063}},
  \href{http://www.arXiv.org/abs/0802.1189}{\texttt{arXiv:0802.1189}}.

\bibitem{Cacciari:2011ma}
\hrefCMSnoop {}{M.~Cacciari, G.~P. Salam, and G.~Soyez, ``{FastJet} user
  manual'',} \textit{ Eur. Phys. J. C} \textbf{ 72} (2012) 1896,
  \href{http://dx.doi.org/10.1140/epjc/s10052-012-1896-2}{\doi{10.1140/epjc/s10052-012-1896-2}},
\href{http://www.arXiv.org/abs/1111.6097}{\texttt{arXiv:1111.6097}}.

\bibitem{CMS:2016lmd}
\hrefCMSnoop {}{{CMS Collaboration}, ``Jet energy scale and resolution in the
  {CMS} experiment in pp collisions at 8 {TeV}'',} \textit{ JINST} \textbf{ 12}
  (2017) P02014,
  \href{http://dx.doi.org/10.1088/1748-0221/12/02/P02014}{\doi{10.1088/1748-0221/12/02/P02014}},
\href{http://www.arXiv.org/abs/1607.03663}{\texttt{arXiv:1607.03663}}.

\bibitem{Sirunyan:2017ezt}
\hrefCMSnoop {}{{CMS Collaboration}, ``Identification of heavy-flavour jets
  with the {CMS} detector in pp collisions at 13 {TeV}'',} \textit{ JINST}
  \textbf{ 13} (2018) P05011,
  \href{http://dx.doi.org/10.1088/1748-0221/13/05/P05011}{\doi{10.1088/1748-0221/13/05/P05011}},
\href{http://www.arXiv.org/abs/1712.07158}{\texttt{arXiv:1712.07158}}.

\bibitem{Sirunyan:2018pgf}
\hrefCMSnoop {}{{CMS Collaboration}, ``Performance of reconstruction and
  identification of $\tau$ leptons decaying to hadrons and $\nu_\tau$ in pp
  collisions at $\sqrt{s}=$ 13 {TeV}'',} \textit{ JINST} \textbf{ 13} (2018)
  P10005,
  \href{http://dx.doi.org/10.1088/1748-0221/13/10/P10005}{\doi{10.1088/1748-0221/13/10/P10005}},
\href{http://www.arXiv.org/abs/1809.02816}{\texttt{arXiv:1809.02816}}.

\bibitem{Alwall:2014hca}
J.~Alwall\hrefCMSnoop {}{ {et~al.}, ``The automated computation of tree-level
  and next-to-leading order differential cross sections, and their matching to
  parton shower simulations'',} \textit{ JHEP} \textbf{ 07} (2014) 079,
  \href{http://dx.doi.org/10.1007/JHEP07(2014)079}{\doi{10.1007/JHEP07(2014)079}},
\href{http://www.arXiv.org/abs/1405.0301}{\texttt{arXiv:1405.0301}}.

\bibitem{Sjostrand:2014zea}
T.~Sj{\"o}strand\hrefCMSnoop {}{ {et~al.}, ``An introduction to {PYTHIA}
  8.2'',} \textit{ Comput. Phys. Commun.} \textbf{ 191} (2015) 159,
  \href{http://dx.doi.org/10.1016/j.cpc.2015.01.024}{\doi{10.1016/j.cpc.2015.01.024}},
\href{http://www.arXiv.org/abs/1410.3012}{\texttt{arXiv:1410.3012}}.

\bibitem{Khachatryan:2015pea}
\hrefCMSnoop {}{{CMS Collaboration}, ``Event generator tunes obtained from
  underlying event and multiparton scattering measurements'',} \textit{ Eur.
  Phys. J. C} \textbf{ 76} (2016) 155,
  \href{http://dx.doi.org/10.1140/epjc/s10052-016-3988-x}{\doi{10.1140/epjc/s10052-016-3988-x}},
\href{http://www.arXiv.org/abs/1512.00815}{\texttt{arXiv:1512.00815}}.

\bibitem{Sirunyan:2019dfx}
\hrefCMSnoop {}{{CMS Collaboration}, ``Extraction and validation of a new set
  of {CMS PYTHIA8} tunes from underlying-event measurements'',} \textit{ Eur.
  Phys. J. C} \textbf{ 80} (2020) 4,
  \href{http://dx.doi.org/10.1140/epjc/s10052-019-7499-4}{\doi{10.1140/epjc/s10052-019-7499-4}},
  \href{http://www.arXiv.org/abs/1903.12179}{\texttt{arXiv:1903.12179}}.

\bibitem{Ball:2014uwa}
\hrefCMSnoop {}{{NNPDF} Collaboration, ``Parton distributions for the {LHC Run
  II}'',} \textit{ JHEP} \textbf{ 04} (2015) 040,
  \href{http://dx.doi.org/10.1007/JHEP04(2015)040}{\doi{10.1007/JHEP04(2015)040}},
\href{http://www.arXiv.org/abs/1410.8849}{\texttt{arXiv:1410.8849}}.

\bibitem{geant4}
\hrefCMSnoop {}{{{GEANT4}} Collaboration, ``{\GEANTfour} --- a simulation
  toolkit'',} \textit{ Nucl. Instrum. Meth. A} \textbf{ 506} (2003) 250,
\href{http://dx.doi.org/10.1016/S0168-9002(03)01368-8}{\doi{10.1016/S0168-9002(03)01368-8}}.

\bibitem{Kalogeropoulos:2018cke}
\hrefCMSnoop {}{A.~Kalogeropoulos and J.~Alwall, ``The {SysCalc} code: A tool
  to derive theoretical systematic uncertainties'',} 2018.
\href{http://www.arXiv.org/abs/1801.08401}{\texttt{arXiv:1801.08401}}.

\bibitem{Beenakker:1999xh}
W.~Beenakker\hrefCMSnoop {}{ {et~al.}, ``Production of charginos, neutralinos,
  and sleptons at hadron colliders'',} \textit{ Phys. Rev. Lett.} \textbf{ 83}
  (1999) 3780,
  \href{http://dx.doi.org/10.1103/PhysRevLett.83.3780}{\doi{10.1103/PhysRevLett.83.3780}},
  \href{http://www.arXiv.org/abs/hep-ph/9906298}{\texttt{arXiv:hep-ph/9906298}}.
  [Erratum: \DOI{10.1103/PhysRevLett.100.029901}].

\bibitem{Nason:2004rx}
\hrefCMSnoop {}{P.~Nason, ``A new method for combining {NLO QCD} with shower
  {M}onte {C}arlo algorithms'',} \textit{ JHEP} \textbf{ 11} (2004) 040,
  \href{http://dx.doi.org/10.1088/1126-6708/2004/11/040}{\doi{10.1088/1126-6708/2004/11/040}},
\href{http://www.arXiv.org/abs/hep-ph/0409146}{\texttt{arXiv:hep-ph/0409146}}.

\bibitem{Frixione:2007vw}
\hrefCMSnoop {}{S.~Frixione, P.~Nason, and C.~Oleari, ``Matching {NLO QCD}
  computations with parton shower simulations: the {POWHEG} method'',} \textit{
  JHEP} \textbf{ 11} (2007) 070,
  \href{http://dx.doi.org/10.1088/1126-6708/2007/11/070}{\doi{10.1088/1126-6708/2007/11/070}},
\href{http://www.arXiv.org/abs/0709.2092}{\texttt{arXiv:0709.2092}}.

\bibitem{Alioli:2010xd}
\hrefCMSnoop {}{S.~Alioli, P.~Nason, C.~Oleari, and E.~Re, ``A general
  framework for implementing {NLO} calculations in shower {M}onte {C}arlo
  programs: the {POWHEG BOX}'',} \textit{ JHEP} \textbf{ 06} (2010) 043,
  \href{http://dx.doi.org/10.1007/JHEP06(2010)043}{\doi{10.1007/JHEP06(2010)043}},
\href{http://www.arXiv.org/abs/1002.2581}{\texttt{arXiv:1002.2581}}.

\bibitem{Re:2010bp}
\hrefCMSnoop {}{E.~Re, ``Single-top {$Wt$}-channel production matched with
  parton showers using the {POWHEG} method'',} \textit{ Eur. Phys. J. C}
  \textbf{ 71} (2011) 1547,
  \href{http://dx.doi.org/10.1140/epjc/s10052-011-1547-z}{\doi{10.1140/epjc/s10052-011-1547-z}},
\href{http://www.arXiv.org/abs/1009.2450}{\texttt{arXiv:1009.2450}}.

\bibitem{Chatrchyan:2013xna}
\hrefCMSnoop {}{{CMS Collaboration}, ``Search for top-squark pair production in
  the single-lepton final state in pp collisions at $\sqrt{s}$ = 8 {TeV}'',}
  \textit{ Eur. Phys. J. C} \textbf{ 73} (2013) 2677,
  \href{http://dx.doi.org/10.1140/epjc/s10052-013-2677-2}{\doi{10.1140/epjc/s10052-013-2677-2}},
\href{http://www.arXiv.org/abs/1308.1586}{\texttt{arXiv:1308.1586}}.

\bibitem{MT2variable}
\hrefCMSnoop {}{C.~G. Lester and D.~J. Summers, ``Measuring masses of
  semi-invisibly decaying particle pairs produced at hadron colliders'',}
  \textit{ Phys. Lett. B} \textbf{ 463} (1999) 99,
  \href{http://dx.doi.org/10.1016/S0370-2693(99)00945-4}{\doi{10.1016/S0370-2693(99)00945-4}},
\href{http://www.arXiv.org/abs/hep-ph/9906349}{\texttt{arXiv:hep-ph/9906349}}.

\bibitem{MT2variable2}
\hrefCMSnoop {}{A.~Barr, C.~Lester, and P.~Stephens, ``{$m_{\mathrm{T2}}$}: the
  truth behind the glamour'',} \textit{ J. Phys. G} \textbf{ 29} (2003) 2343,
  \href{http://dx.doi.org/10.1088/0954-3899/29/10/304}{\doi{10.1088/0954-3899/29/10/304}},
\href{http://www.arXiv.org/abs/hep-ph/0304226}{\texttt{arXiv:hep-ph/0304226}}.

\bibitem{MT2variable3}
\hrefCMSnoop {}{C.~G. Lester and B.~Nachman, ``Bisection-based asymmetric
  {$m_{\mathrm{T2}}$} computation: a higher precision calculator than existing
  symmetric methods'',} \textit{ JHEP} \textbf{ 03} (2015) 100,
  \href{http://dx.doi.org/10.1007/JHEP03(2015)100}{\doi{10.1007/JHEP03(2015)100}},
\href{http://www.arXiv.org/abs/1411.4312}{\texttt{arXiv:1411.4312}}.

\bibitem{CMS-LUM-17-003}
\hrefCMSnoop {}{{CMS Collaboration}, ``Precision luminosity measurement in
  proton-proton collisions at $\sqrt{s} =$ 13 {TeV} in 2015 and 2016 at
  {CMS}'',} \textit{ Eur. Phys. J. C} \textbf{ 81} (2021) 800,
  \href{http://dx.doi.org/10.1140/epjc/s10052-021-09538-2}{\doi{10.1140/epjc/s10052-021-09538-2}},
  \href{http://www.arXiv.org/abs/2104.01927}{\texttt{arXiv:2104.01927}}.

\bibitem{CMS-PAS-LUM-17-004}
\href {http://cds.cern.ch/record/2621960}{{CMS Collaboration}, ``{CMS}
  luminosity measurement for the 2017 data-taking period at {$\sqrt{s} =
  13~\mathrm{TeV}$}'',} CMS Physics Analysis Summary CMS-PAS-LUM-17-004, 2018.

\bibitem{CMS-PAS-LUM-18-002}
\href {http://cds.cern.ch/record/2676164}{{CMS Collaboration}, ``{CMS}
  luminosity measurement for the 2018 data-taking period at {$\sqrt{s} =
  13~\mathrm{TeV}$}'',} CMS Physics Analysis Summary CMS-PAS-LUM-18-002, 2019.

\bibitem{Simp}
\hrefCMSnoop {}{{CMS Collaboration}, ``Interpretation of searches for
  supersymmetry with simplified models'',} \textit{ Phys. Rev. D} \textbf{ 88}
  (2013) 052017,
  \href{http://dx.doi.org/10.1103/PhysRevD.88.052017}{\doi{10.1103/PhysRevD.88.052017}},
\href{http://www.arXiv.org/abs/1301.2175}{\texttt{arXiv:1301.2175}}.

\bibitem{Junk:1999kv}
\hrefCMSnoop {}{T.~Junk, ``Confidence level computation for combining searches
  with small statistics'',} \textit{ Nucl. Instrum. Meth. A} \textbf{ 434}
  (1999) 435,
  \href{http://dx.doi.org/10.1016/S0168-9002(99)00498-2}{\doi{10.1016/S0168-9002(99)00498-2}},
\href{http://www.arXiv.org/abs/hep-ex/9902006}{\texttt{arXiv:hep-ex/9902006}}.

\bibitem{Read:2002hq}
\hrefCMSnoop {}{A.~L. Read, ``Presentation of search results: the
  {$\text{CL}_\text{s}$} technique'',} \textit{ J. Phys. G} \textbf{ 28} (2002)
  2693,
\href{http://dx.doi.org/10.1088/0954-3899/28/10/313}{\doi{10.1088/0954-3899/28/10/313}}.

\bibitem{CMS-NOTE-2011-005}
\href {https://cds.cern.ch/record/1379837}{{ATLAS and CMS Collaborations, and
  the LHC Higgs Combination Group}, ``Procedure for the {LHC} {Higgs} boson
  search combination in {Summer} 2011'',} Technical Report CMS-NOTE-2011-005,
  ATL-PHYS-PUB-2011-11, 2011.

\bibitem{Cowan:2010js}
\hrefCMSnoop {}{G.~Cowan, K.~Cranmer, E.~Gross, and O.~Vitells, ``Asymptotic
  formulae for likelihood-based tests of new physics'',} \textit{ Eur. Phys. J.
  C} \textbf{ 71} (2011) 1554,
  \href{http://dx.doi.org/10.1140/epjc/s10052-011-1554-0}{\doi{10.1140/epjc/s10052-011-1554-0}},
  \href{http://www.arXiv.org/abs/1007.1727}{\texttt{arXiv:1007.1727}}.
  [Erratum: \DOI{10.1140/epjc/s10052-013-2501-z}].

\end{thebibliography}\endgroup
\cleardoublepage \appendix\section{The CMS Collaboration \label{app:collab}}\begin{sloppypar}\hyphenpenalty=5000\widowpenalty=500\clubpenalty=5000
\cmsinstitute{Yerevan Physics Institute, Yerevan, Armenia}
{\tolerance=6000
A.~Tumasyan\cmsAuthorMark{1}\cmsorcid{0009-0000-0684-6742}
\par}
\cmsinstitute{Institut f\"{u}r Hochenergiephysik, Vienna, Austria}
{\tolerance=6000
W.~Adam\cmsorcid{0000-0001-9099-4341}, J.W.~Andrejkovic, T.~Bergauer\cmsorcid{0000-0002-5786-0293}, S.~Chatterjee\cmsorcid{0000-0003-2660-0349}, K.~Damanakis\cmsorcid{0000-0001-5389-2872}, M.~Dragicevic\cmsorcid{0000-0003-1967-6783}, A.~Escalante~Del~Valle\cmsorcid{0000-0002-9702-6359}, R.~Fr\"{u}hwirth\cmsAuthorMark{2}\cmsorcid{0000-0002-0054-3369}, M.~Jeitler\cmsAuthorMark{2}\cmsorcid{0000-0002-5141-9560}, N.~Krammer\cmsorcid{0000-0002-0548-0985}, L.~Lechner\cmsorcid{0000-0002-3065-1141}, D.~Liko\cmsorcid{0000-0002-3380-473X}, I.~Mikulec\cmsorcid{0000-0003-0385-2746}, P.~Paulitsch, F.M.~Pitters, J.~Schieck\cmsAuthorMark{2}\cmsorcid{0000-0002-1058-8093}, R.~Sch\"{o}fbeck\cmsorcid{0000-0002-2332-8784}, D.~Schwarz\cmsorcid{0000-0002-3821-7331}, S.~Templ\cmsorcid{0000-0003-3137-5692}, W.~Waltenberger\cmsorcid{0000-0002-6215-7228}, C.-E.~Wulz\cmsAuthorMark{2}\cmsorcid{0000-0001-9226-5812}
% ...
\par}
\vskip\cmsinstskip
\dag:~Deceased\\
$^{1}$Also at Yerevan State University, Yerevan, Armenia\\
$^{2}$Also at TU Wien, Vienna, Austria\\
% ...
\end{sloppypar}
\end{document}